\documentstyle[12pt,preprint,aps,epsfig]{revtex}
\tighten
\newcommand{\beq}{\begin{eqnarray}}
\newcommand{\eeq}{\end{eqnarray} }
\newcommand{\bsga}{B \to X_s \gamma }
\newcommand{\bsgg}{B \to X_s \gamma \gamma }
\newcommand{\tab}[1]{Table \ref{#1}}

\newcommand{\non}{\nonumber\\ }
\newcommand{\rgg}{R_{\gamma \gamma}}
\newcommand{\rg}{R_{\gamma} }
\newcommand{\mhp}{ m_{H^\pm} }
\def\gsim{ {\ \lower-1.2pt\vbox{\hbox{\rlap{$>$}\lower5pt\vbox{\hbox{$\sim$}}}}\ } }

\title{ {\bf The $ B \to X_s \gamma \gamma $ decay in the standard model and
the general two-Higgs-doublet model}}
\author{ Junjie Cao \\
{\footnotesize Department of Physics, Henan Normal University, Xinxiang, Henan,
453002, People's Republic of China}  \\
\vspace{.5cm}
  Zhenjun Xiao\\
{\footnotesize CCAST(World Laboratory), P.O.Box 8730, Beijing 100080,
People's Republic of China }\\
{\footnotesize Department of Physics, Henan Normal University, Xinxiang, Henan,
453002, People's Republic of China}  \\
{\footnotesize Department of Physics, Peking University, Beijing, 100871,
People's Republic of China} \\
\vspace{0.5cm}
  Gongru Lu\\
{\footnotesize  CCAST(World Laboratory), P.O.Box 8730, Beijing 100080,
People's Republic of China }\\
{\footnotesize  Department of Physics, Henan Normal University, Xinxiang, Henan,
453002, People's Republic of China}  \\ }
\date{\today}
\begin{document}
\maketitle
\begin{abstract}
Based on the low-energy effective Hamiltonian, we calculate the
new physics corrections to the branching ratio and the
differential distributions of the rare decay $B \to X_s \gamma
\gamma$ induced by the new gluonic and electroweak charged-Higgs
penguin diagrams in  the general two-Higgs-doublet model with the
restriction $\lambda_{ij}^{U,D}=0$ for $i\neq j$.
Within the considered parameter space, we see the following:
(a) the standard model predictions of $ {\cal B}( B \to X_s \gamma
\gamma) $ and $A_{FB}$ have a moderate $m_s$ dependence;
(b)  in model III , the prediction of the branching ratio $ {\cal B}( B
\to X_s \gamma \gamma) $ ranges from one third to three times of the
standard model prediction, but is highly correlated with that of $ {\cal
B}( B \to X_s \gamma )$;
(c) the new physics enhancement to the branching ratio $ {\cal B}( B
\to X_s \gamma \gamma) $ in model II can be as large as $(30-50)\%$;
(d) the contribution from 1PR diagrams is
dominant and hence four normalized differential  distributions are
insensitive to the variation of scale $\mu $ and possible new
physics corrections;
(e) due to the smallness of its decay rate and the long-distance background,
 the $B \to X_s \gamma \gamma$ decay is not a better process in
 detecting new physics than the $B \to X_s \gamma$ decay.
\end{abstract}

\vspace{1cm}
\noindent
PACS numbers:~13.25.Hw,~12.15.Ji,~12.40.Bx, 12.60.Fr

\newpage
\section{Introduction}

The flavor changing neutral current (FCNC) inclusive transition $ B \to X_s
\gamma $ has been a subject of great interest during the past
few years. The basic theoretical framework is the standard
model (SM)  at scales $m_w$ or $m_t$. QCD short distance
corrections \cite{o1} are then incorporated via the
renormalization group technique to yield the low energy effective
Hamiltonian valid at scale $O(m_b) $ which is relevant for B decay
processes\cite{bbl96,bf97}.

As is well known, the rare decay $ B  \to X_s \gamma $ is theoretically
very clear, measured at CLEO \cite{cleo95,cleo99} with
increasing accuracy and  in remarkable agreement with theoretical
estimates. Great progress in both the theoretical calculation
\cite{kagan99} and the experimental measurement \cite{cleo99} for
$B \to X_s \gamma$ decay enable
one to constrain the new physics models, such as the two-Higgs-doublet
model (2HDM) \cite{2hdm}, the minimal supersymmetric standard
model (MSSM) \cite{misiak97} and the Technicolor models \cite{xiao96}.
In light of these developments, it is natural to consider other
inclusive channels which as  a whole may separate out contributions
from various operators in the effective Hamiltonian.

The rare flavor changing neutral current inclusive transition $ B
\to X_s \gamma \gamma$, which is expected to be $\sim 10^{-3} $
smaller in its branching ratio relative to $B\to X_s \gamma$, has
attracted new attention in view of the planned experiments at KEK
and SLAC B-factories and existing accelerators, which may test
branching ratios as low as $10^{-8}$. Just like the decay $ B \to
X_s \gamma $, the decay $\bsgg$ is relatively clean after some
proper precaution to take out effects due to strong resonance, for
instance, the effect of $ \eta_c $ at its peak to the two photon
spectrum \cite{o3}. The studies about these rare decay modes will
then provide further opportunities in testing the whole technology
of weak decays, or better yet in providing some clues of new
physics.

There have been some theoretical investigations for the process $
\bsgg$, which corresponds at the quark level to the transition $
b\to s \gamma \gamma$. Theoretical calculations in the SM were
firstly done on the basis of pure electroweak theory \cite{o4,o5}
and subsequently improved to include the leading order (LO) QCD
effects \cite{o6,o7}. Nearly all the studies of  $\bsgg$ and
$\bsga$ decays are based on the free  decay of $b$ quark and can
be justified from the heavy quark effective theory (HQET).
According to the argument in Ref.\cite{o8}, the HQET corrections
to these rare decay modes are strongly suppressed by powers of
$(\Lambda/m_b)^2$. Furthermore, Choudhury and Yao \cite{o9} have
shown that even when$ (\Lambda/m_c)^2 $ and $ (\Lambda/m_c)^4 $
corrections are included, the overall HQET corrections to the free
quark results of $\bsgg$ are only a few percent and can be safely
neglected. In this paper, we will present our results at free
quark level.

Recently, we estimated the new physics contributions to the
inclusive decay $b \to q g^* \to q \overline{q'} q'$ with $q \in
\{ d,s \}$ and $q' \in \{ u,d,s\}$  in the model III: the
third-type two-Higgs-doublet model \cite{hou92,atwood97}, and
found that the new physics contributions can be significant
\cite{xiao2001}. In this paper, we calculate the new physics
corrections to the branching ratio and differential distributions
of the rare decay $B \to X_s \gamma \gamma$ induced by the new
gluonic and electroweak charged-Higgs penguin diagrams in the
model III.

The plan of this paper is as follows:  in the next section, we
give a brief review of the process $ B \to X_s \gamma \gamma $ in
the SM at LO approximation, present the relevant formulae and
discuss some general characteristics of this process. In Sec. III,
we will investigate in detail this rare decay in the model III and
present some numerical results. In the last section, concluding
remarks are added.

\section{$B \to X_s \gamma \gamma$ in the SM}

By the use of an extension of Low's low energy theorem
\cite{o6,low}, or alternatively, by applying the equation of
motions \cite{o10}, the most general effective Hamiltonian that
describes radiative $ b \to s $ decays with up to three emitted
gluons or photons is given by \footnote{Strictly speaking,
operators $ (\bar{s}_\alpha \gamma^{\mu} L u_\alpha)
(\bar{u}_\beta \gamma_{\mu} L b_\beta )$ and $ (\bar{s}_\alpha
\gamma^{\mu} L u_\beta) (\bar{u}_\beta \gamma_{\mu} L b_\alpha) $
should be added. Here we neglect them for their small
coefficients.}

\beq
 H_{eff}&=&-\frac{4 G_F}{\sqrt{2}} V_{tb} V_{ts}^* \sum_{i=1}^{8}
 C_i(\mu) O_{i}~, \label{a1}
\eeq where the current-current, penguin, electroweak- and
chromo-magnetic dipole operators are of the form
\beq
O_1&=&(\bar{s}_\alpha \gamma^{\mu} L c_\beta)(\bar{c}_\beta
\gamma_{\mu} L b_\alpha) ~, \non
 O_2&=&(\bar{s}_\alpha
\gamma^{\mu} L c_\alpha)(\bar{c}_\beta \gamma_{\mu} L b_\beta)~,\nonumber \\
 O_3&=&(\bar{s}_\alpha
\gamma^{\mu} L b_\alpha) \sum_{q=u \cdots b}(\bar{q}_\beta
\gamma_{\mu} L q_\beta)~, \nonumber \\
 O_4&=&(\bar{s}_\alpha
\gamma^{\mu} L b_\beta) \sum_{q=u \cdots b}(\bar{q}_\beta
\gamma_{\mu} L q_\alpha)~, \nonumber \\
 O_5&=&(\bar{s}_\alpha
\gamma^{\mu} L b_\alpha) \sum_{q=u \cdots b}(\bar{q}_\beta
\gamma_{\mu} R q_\beta)~, \nonumber \\
 O_6&=&(\bar{s}_\alpha
\gamma^{\mu} L b_\beta) \sum_{q=u \cdots b}(\bar{q}_\beta
\gamma_{\mu} R q_\alpha)~, \nonumber \\
O_{7}&=&\frac{e}{16 \pi^2} \bar{s}_{\alpha} \sigma^{\mu \nu} ( m_b R + m_s L)
b_{\alpha} F_{ \mu \nu}~,   \nonumber  \\
O_{8}&=&\frac{g_s}{16
\pi^2} \bar{s}_{\alpha} \sigma^{\mu \nu} ( m_b R + m_s
L)\lambda_{\alpha \beta}^{a} b_{\beta} G_{ \mu \nu}^a~. \label{operator}
\eeq
where $\alpha$ and $\beta$ are color indices, $a = $1,
$\cdots$,8 labels $SU(3)_C$ generators, and $L,R$=$(1  \mp\gamma_5)/2$, while
$F_{\mu \nu}$ and $G_{\mu \nu}^a $ denote the QED
and QCD field strength tensors, respectively.

The Wilson coefficients $ C_{i}(\mu)$ in Eq.(\ref{a1}) are process
independent and their renormalization is determined only by the
basis of operators $ O_{i} $. In our calculation of the leading
order matrix elements of $ b \to s \gamma \gamma $  shown in
Fig.\ref{fig1}, we find no regulation scheme dependence enters
through the new class of penguin diagrams with two external
photon. This means that it is suffice to  use LO
regularization-scheme-independent Wilson coefficients
$C_{1,\cdots,6}$, $C_{7\gamma}$ and $C_{8g}$\cite{o11} and we need
not consider the matrix elements due to the insertion of $O_5$ and
$O_6$ into the one-photon penguin diagrams. In Table \ref{table1},
we present the numerical values of the Wilson coefficients at the
scale $\mu =O(m_b)$ \footnote{ These Wilson coefficients are
numerically well consistent with those given in Ref.\cite{o7}. }.

At the leading order, the amplitude for the decay $b(p) \to s(p^{\prime}) +
\gamma(k_1)+\gamma (k_2) $ can be expressed as\cite{o7}
\beq
A&=&-\frac{i e^2 G_F}{\sqrt{2} \pi^2} V_{t b} V_{t s}^*
\bar{u}_{s}(p^\prime) \cdot \left [ F_2 W_2^{\mu \nu}+F_5 (m_b W_{5, b}^{\mu
\nu}\; R+ m_s W^{\mu \nu}_{5, s}\; L)\right. \non
&& \left. + F_7 W_7^{\mu \nu} \right ] u_b(p)
\epsilon_{\mu}(k_1) \epsilon_{\nu} (k_2)~,
\eeq
with
\beq
F_2&=&\left [ N_c C_1(\mu) +C_2 (\mu) \right ] Q_u^2 \kappa_c \nonumber \\
& &+C_3(\mu)  \left \{  N_c \left [Q_d^2 (\kappa_d+\kappa_s+\kappa_b)+Q_u^2
(\kappa_u+\kappa_c) \right ]+Q_d^2 (\kappa_s+\kappa_b) \right \} \nonumber  \\
& & +C_4(\mu) \left \{ Q_d^2 \left [(\kappa_d+\kappa_s+\kappa_b)+Q_u^2
(\kappa_u+\kappa_c) \right ]+N_c Q_d^2 (\kappa_s+\kappa_b) \right \} \nonumber
\\ & & -\left [N_c C_5(\mu)+C_6(\mu) \right ]  \left [Q_d^2
(\kappa_d+\kappa_s+\kappa_b)+Q_u^2 (\kappa_u+\kappa_c)\right  ]~,  \\
F_5&=&\left [C_5 (\mu) +N_c C_6 (\mu) \right ] Q_d^2~,    \\
F_7&=&C_7 (\mu) Q_d~,    \\
W_2^{\mu \nu} &=&- i \left \{  \frac{1}{k_1 \cdot k_2} \left [ k_1^{\nu}
\epsilon_{\mu \rho\sigma \lambda} \gamma^{\rho}
k_1^{\sigma} k_2^{\lambda}
- k_2^{\mu} \epsilon_{\nu \rho \sigma \lambda} \gamma^{\rho}
 k_1^{\sigma} k_2^{\lambda} \right ]+ \epsilon_{\mu \nu \rho  \lambda}
  \gamma^{\rho} (k_2-k_1)^{\lambda}  \right \} L ~,    \\
W_{5, q}^{\mu \nu}&=& \frac{1}{m_q^2} \left ( - i \epsilon_{\mu \nu  \lambda \sigma }
    k_1^{\lambda} k_2^{\sigma} \gamma_5+ k_1 \cdot k_2 g^{\mu\nu}
    - k_1^{\nu} k_2^{\mu} \right ) \left (1- 2 \kappa_{q} \right )+4 \left ( g^{\mu\nu}
    -\frac{k_1^{\nu} k_2^{\mu}}{k_1 \cdot k_2} \right ) \kappa_q~,  \\
W_7^{\mu \nu}&=&\frac{1}{2} \left [-\frac{1}{2 p \cdot k_2} \not\!
k_1 \gamma^{\mu} ( m_b R+ m_s L) ( \not\! p- \not\! k_2+m_b)
\gamma^{\nu}\right.  \nonumber \\ & & \left. +\frac{1}{2
p^{\prime} \cdot k_2} \gamma^{\nu} (\not\! p- \not\! k_1+m_s)
\not\! k_1 \gamma^{\mu} (m_b R+m_s L) \right ]~,
\eeq
where $k_1$ and $k_2$ are four-momenta of the emitted photons, $N_c=3$
denotes the number of colors, $Q_u=2/3$ and $ Q_d=-1/3 $ are  the up-type
and down-type quark electric charges, and the factor $\kappa_q$ is
defined as
\beq \kappa_q &=& \frac{1}{2}+\frac{1}{z_q} \int_0^1
\frac{d x}{x} \ln \left (1-z_q x + z_q^2 x^2 \right )  \nonumber
\\ & =& \left \{  \begin{array}{ll}  \frac{1}{2}-\frac{2}{z_q}
\left (\arctan \sqrt{\frac{z_q}{4-z_q}}  \right )^2  &   {\rm if
}\ \  z_q<4~, \\ \frac{1}{2}+\frac{1}{z_q} \left
[-\frac{\pi^2}{2}+2 \left ( \ln \frac{\sqrt{z_q}+\sqrt{z_q-4}}{2}
\right ) ^2- 2 i  \pi \ln \left (\frac{\sqrt{z_q}+\sqrt{z_q-4}}{2}
\right ) \right ] \ \ \  & {\rm otherwise}~. \end{array}   \right.
\label{eq:kappa}
\eeq
Here $z_q = 2 k_1 \cdot k_2/m_q^2$. The
masses appearing in $W_7^{\mu \nu} $ arise from operator $O_7$
while those in $ W_{5,q}^{\mu \nu} $ are internal quark masses in
the loops (see Fig.1). In our numerical analysis, we take all
these quark masses as current mass. The factor $\kappa_q$ in
Eq.(\ref{eq:kappa}) is the loop integral function \cite{ag91} and
its absorption part reflects intermediate quark threshold effect.
In getting the expressions of $ W_i^{\mu \nu} $ that are essential
to obtain the matrix elements of $ b \to s \gamma \gamma $ decay
presented in \cite{o6}, we used the following identities: \beq
\gamma_{\mu} \gamma_{\lambda} \gamma_{\sigma} &=& g_{\sigma
    \lambda} \gamma_{\mu}+ g_{\mu \lambda} \gamma_{\sigma}-g_{\mu
    \sigma} \gamma_{\lambda}+  i \epsilon_{\mu \rho \sigma \lambda}
    \gamma^{\rho} \gamma_5~,   \\
\gamma_{\mu_1} \gamma_{\mu_2} \gamma_{\mu_3} \gamma_{\mu_4}&=& i \epsilon_{\mu_1
\mu_2 \mu_3 \mu_4} \gamma_5- g_{\mu_3 \mu_4} g_{\mu_1
\mu_2}+g_{\mu_2 \mu_4} g_{\mu_1 \mu_3}-g_{\mu_2 \mu_3} g_{\mu_1
\mu_4} +g_{\mu_3 \mu_4} \gamma_{\mu_1} \gamma_{ \mu_2}\nonumber \\
& &-g_{\mu_2 \mu_4} \gamma_{\mu_1} \gamma_{ \mu_3}+g_{\mu_2 \mu_3}
\gamma_{\mu_1} \gamma_{ \mu_4}+g_{\mu_1 \mu_4} \gamma_{\mu_2}
\gamma_{ \mu_3}  -g_{\mu_1 \mu_3} \gamma_{\mu_2} \gamma_{
\mu_4}+g_{\mu_1 \mu_2} \gamma_{\mu_3} \gamma_{ \mu_4}~.
\eeq

The square amplitude summed over spins and polarizations  is then
given by
\beq
|A|^2&=& \frac{1}{4} \left (\frac{e^2 G_F}{\sqrt{2}
\pi^2} V_{t b} V_{t s}^* \right )^2 m_b^4 \left \{  |F_2|^2
A_{22}+|F_5|^2 A_{55}+|F_7|^2 A_{77} + 2 Re[F_7 F_2^*] A_{27}
\right.   \non &&\left. +2 Re[F_5 F_2^* (1- 2\kappa_s)] A_{2 5}^s+
2 Re[F_7 F_5^*] A_{5 7} +2 Re[F_5 F_2^* (1-2 \kappa_b)]A_{2 5}^b
\right \}~, \label{eq:amps}
\eeq
where the quantities $ A_{ij} $
denote the contractions between the tensor $W_i^{\mu \nu *}$ and $
W_j^{\mu \nu}$, and the first two terms and the third term of
$|A|^2$ arise from one-particle irreducible (1PI) diagrams and
one-particle reducible (1PR) diagrams, respectively. In order to
give the explicit expressions of $ A_{ij} $, we introduced the
following notations,
\beq
s&=&\frac{2 k_1 \cdot k_2}{m_b^2},\ \ \ \
t=\frac{2 p \cdot k_2}{m_b^2},\ \ \ \
u=\frac{2 p \cdot k_1}{m_b^2}, \ \ \ \ \
rho_1=\frac{m_{s}^2}{m_b^2}~, \non
\rho&=&1-u-t+s=\frac{p^{\prime 2}}{m_b^2},\ \ \ \
\bar{s}=\frac{s}{1-\rho},\ \ \ \ \bar{t}=\frac{t}{1-\rho},\ \ \ \
\bar{u}=\frac{u}{1-\rho}~,
\eeq
where $ m_{s} $ is $s $ quark mass
appearing in $ W_{5, 7}^{\mu \nu} $ and $p^{\prime} $ is the
momentum of the outgoing $ s $ quark. In this framework,
coefficients $ A_{i j}$ are then given by
\beq
A_{22} &=& 2 [(1-\rho)^2-(1+\rho) s]~,  \label{eq:a22}  \\
A_{55} &=&\left \{ 16 |\kappa_b |^2 + |(1- 2 \kappa_b) s
+4 \kappa_b|^2+\rho_1 [ 16 |\kappa_s|^2+ | (1- 2 \kappa_s) \frac{s}{\rho_1}
+4 \kappa_s |^2 ] \right \} (1- s+\rho)  \non
& & +16 Re \left \{ 8 \rho_1
\kappa_b \kappa_s^*+s [\kappa_b-2 (1+\rho_1) \kappa_b \kappa_s^*+
\rho_1 \kappa_s^*]\right \}~,  \label{eq:a55}   \\
A_{25}^{b,s}&=&\pm s (1-\rho \mp s)~,  \label{eq:a25}  \\
A_{27}&=& - \left [2 (1+\rho_1) s+ \frac{(\rho+\rho_1) s^2
+ 2 (\rho_1-\rho) t s}{t (s-t)}+\frac{(\rho+\rho_1) s^2
+ 2 (\rho_1-\rho) u s}{u (s-u)} \right ] ~,  \label{eq:a27}  \\
A_{57}&=& Re \left \{ 8 (\kappa_b+
\rho_1 \kappa_s) s- \left [  4 \rho_1 (\kappa_b+ \kappa_s)+s \left
[(1- 2 \kappa_s)+ \rho_1 (1- 2 \kappa_b) \right ] \right ]\right.
\non & & \left. \times \left [ \frac{s^2}{t (s-t)}+ \frac{s^2}{u
(s-u)} \right]-4 \left [(\kappa_b+\rho_1 \kappa_s) (\rho - \rho_1)
(\frac{s^2}{t (s-t)}+ \frac{s^2}{u (s-u)}) \right ] \right  \}~,
\label{eq:a57}   \\
A_{77}&=&(1+\rho_1) \left [(1-\rho) A_{77}^1+
A_{77}^2 \right ]+A_{77}^3 ~, \label{eq:a77}  \\ A_{77}^1&=&
\frac{1}{\bar{t}} \left [1+\bar{u} +\frac{2 \bar{u}
(\bar{u}-2)}{(1-\bar{u})} \bar{t} +\frac{2 \bar{u}-1}{1-\bar{u}}
\bar{t}^2 \right ]+(\bar{t} \leftrightarrow \bar{u} )~, \label{eq:a771}  \\
A_{77}^2&=& \frac{-2}{\bar{t}^2} \left [
1-\frac{1+\rho_1}{1-\bar{u}} \bar{t}+\frac{\rho_1}{(1-\bar{u})^2}
\bar{t}^2 \right ] \nonumber  \\ &&+(\rho_1-\rho) \left ( \frac{2
\bar{s}}{\bar{t} (1-\bar{u})}-\frac{\bar{s}
\bar{u}}{(1-\bar{u})^2}-\frac{\bar{u}^2}{(1-\bar{u})^2} \right )
+(\bar{t} \leftrightarrow \bar{u} )   ~,  \label{eq:a772}  \\
A_{77}^3&=&-2 \frac{s}{\bar{t} \bar{u}} t \left \{  (1+\rho_1) (2+
\bar{u} \bar{t})+\frac{\rho_1}{1- \rho} \left [ 1- \frac{2 (1+
\rho_1) -\bar{t} \bar{u}}{(1-\bar{t}) (1-\bar{u})} \right ]
\bar{s} \right \}  \nonumber  \\ & & + \frac{2 s}{t u}  (\rho_1-
\rho) \frac{- (1+ \rho_1) (2 s +\bar{s}) -\rho_1 ( \bar{s}^2- 2 \
\bar{t}\  \bar{u})+2 (2-\rho) (1+\rho_1) \bar{t}\
\bar{u}}{(1-\bar{u})(1-\bar{t})}~, \label{eq:a773}
\eeq
Unlike Ref.\cite{o7}, we
here distinguish different $ s $ quark contributions which enter
in the amplitude via $ \rho $ and $ \rho_1 $ respectively. This
is crucial to our results, as described below. Note
that in the $A_{77} $ term of the square amplitude, there exists
infrared (IR) divergence when one integrates the amplitude over
the physical phase space. This divergence can be canceled out when
one considers $ O(\alpha_e) $ virtual corrections to the $ b \to s
\gamma $ amplitude \cite{apqcd}. In order to calculate the
physical rate of interest that is free of divergence, we have to
impose a cut on the energy of each photon, which will naturally
correspond to the experimental cut imposed on the minimum energy
of detectable photons.

In numerical calculations, we use the input parameters listed in
\tab{table2} and take the following cuts: \beq E_{\gamma}> 100\
MeV~, \ \ \ \ E_{s}> 600\  MeV~,  \ \ \ \ \theta>\frac{\pi}{9}~,
\label{eq:cuts} \eeq where $ E_{\gamma} $ and $ E_s $ denote the
energy of photon and that of of the outgoing mesons respectively,
and $\theta $ is the angle between any two outgoing particles. The
first constraint is required to avoid $ IR $ divergence while the
last constraint is to exclude photons that are emitted too close
to each other or to the outgoing s quark.  For the mass of $ s  $
quark, cares must be taken. In principle, constituent mass should
be used in the phase space integration, while the masses appeared
in $ W_7^{\mu \nu} $ and $ W_5^{\mu \nu} $ should be the current
mass. This means that $\rho $ and $ \rho_1 $ appeared in
Eqs.(\ref{eq:a22}-\ref{eq:a773}) should be
\beq
\rho=\frac{m_s^2\
(constituent)}{m_b^2} \simeq \frac{m_K^2}{m_b^2}~, \ \ \ \
\rho_1=\frac{m_s^2 \ (current)}{m_b^2}~.
\eeq

Following Ref.\cite{o12}, the branching ratio of $\bsgg$ decay can
be written as
\beq
{\cal B} ( B \to X_s \gamma \gamma) \simeq
\left [\frac{\Gamma (b \to s \gamma \gamma)}{ \Gamma (b \to c l
\nu_{l})} \right ]^{th} \times {\cal B}  ( B \to X_c l
\nu_l)^{expt}~. \label{eq:brbsgg}
\eeq
Using the input parameters presented in Table \ref{table2}, setting
$\mu =m_b$ and taking the cuts given in Eq.(\ref{eq:cuts}), we find
numerically that
\beq
{\cal B}( \bsgg ) \approx  4.6 \times 10^{-7}~,\ \
A_{FB}(\bsgg)=0.79~,
\eeq
where $A_{FB}$ is the forward-backward asymmetry of the $\bsgg$ decay.

If we use a common $ s $ quark mass $m_s=0.5$ GeV in the numerical
calculations as Ref.\cite{o7} did, we find that
\beq
{\cal B}(\bsgg)= 3.8 \times 10^{-7}\, , \ \ \ A_{FB}(\bsgg)=0.76\, ,
\eeq
which agree well with the results presented in Ref.\cite{o7},
where ${\cal B}( B \to X_s \gamma \gamma )= 3.7 \times 10^{-7}$
and $A_{FB}=0.78$. For $ m_s=0.15$ GeV, however, we find that
\beq
{\cal B}( \bsgg )= 5.1 \times 10^{-7}\, , \ \ \
A_{FB}(\bsgg)=0.81~.
\eeq
From above numerical results, we find that the $m_s$-dependence is weak
for $A_{FB}$, but relatively strong for the branching ratio.

Among all the contributions to the decay rate,  the contribution
from 1PR diagrams is predominant, larger than $97\%$ of the total,
which is due to the cancellation between $C_1 $ and $C_2$ and the
QCD enhancement of $|C_7|$. Numerical results also show that the
branching ratio is sensitive to the cuts we have imposed. For
example, if we demand $\theta $ in Eq.(\ref{eq:cuts}) larger than
$\pi/6  $, the branching ratio will reduce to $ 3.6 \times
10^{-7}$. Moreover, because of the scale dependence of Wilson
coefficients, there exists a $\sim 25 \%$ theoretical uncertainty
at leading order approximation, as is the case for $ B \to X_s
\gamma $.

In Fig.\ref{fig2}, we present the normalized differential
distribution $ (1/\Gamma) d\Gamma/d s$ versus $ s $ ( solid curve ).
Comparing with the corresponding result given in \cite{o5}, we find that
the peak due to $c\bar{c}$ threshold effect is smeared and the average invariant mass
of the two photons is lowered when QCD corrections are added. In
Figs.(\ref{fig3}-\ref{fig5}) , the normalized differential distribution
$ (1/\Gamma) d\Gamma/d \cos\theta_{\gamma \gamma} $ versus
$\cos\theta_{\gamma \gamma} $ (solid curve in Fig.\ref{fig3}) and
the spectrum of the two photons, defined as the photon with lower
energy (solid curve in Fig.\ref{fig4}) and the photon with  higher
energy (solid curve in Fig.\ref{fig5}), are plotted.

From these four figures, one can understand the kinematics of the
process as follows: {\it an energetic $ s $ quark with mean energy
around $ 2.0 GeV $ tends to be emitted, compensated by the harder
of the two photons; while the less energetic photon tends to go in
the direction of  the $ s $ quark. This topology is typical of a
bremsstrahlung event of the $ s $ quark. } Such kinematics are
very useful for us to separate the short-distance (SD) signal from
the long-distance (LD) background, which may come from the channel
$ B \to X_s\ \eta\ (\eta^{\prime}) \to X_s \gamma \gamma $
\cite{o3,o5}. For example, by demanding $s \leq 0.3 $ and $\cos
\theta_{\gamma \gamma} \leq 0 $, about $75 \%$ of the signal is
remained while almost all the LD background are removed.

In Figs.(\ref{fig2}-\ref{fig5}), the contribution from 1PI
diagrams (dashed curve) and that from the interference between 1PI
and 1PR (dot-dashed curve) are also plotted. From these figures,
we can see that except in some marginal area, the contributions
from 1PI diagrams and the interference are much smaller than those
coming from 1PR diagrams. Taking this in mind, one can infer that
though the four normalized differential distributions are
calculated at leading order approximation, they are  insensitive
to the variation of scale. The reason is that, to a good
approximation, a common factor $ |C_7 (\mu)|^2 $ can be extracted
from both $ \Gamma $ and $ d \Gamma$ and will disappear in the
ratio of them. This feature becomes more evident when the cuts of
$s \leq 0.3 $ and $\cos \theta_{\gamma \gamma} \leq 0 $ are
imposed, which removes nearly all the contributions of $ O_1 \sim
O_6 $ (see Fig.\ref{fig2}). In fact, a more general statement is
that: as long as the new physics effects beyond the SM appear only
in the matching of the Wilson coefficients of the standard
effective operator basis, which is so in the MSSM \cite{o12}, and
$|C_7|$ is not suppressed as required by $ b \to s \gamma $, these
four observable distributions are insensitive to the variation of
scale $\mu$ and the new physics.

\section{$B \to X_s \gamma \gamma $ In Model III}

\subsection{Model description}

The two-Higgs-doublet model (2HDM) \cite{2hdm} is the simplest
extension of the SM . During past years, the models I and II have
been studied extensively in literature and tested experimentally
and the model II has been very popular since it is the building
block of the MSSM. In Ref.\cite{o5}, the authors studied the
$\bsgg$ decay in the SM and models I and II in the basis of pure
electroweak theory \cite{o4}, and found that the branching ratios
in models I and II can be appreciably different from that in the
SM\cite{o5}. In this paper, we focus on estimating the new physics
effects on the $\bsgg$ decay in the framework of the third type of
2HDM, usually known as the model III \cite{atwood97}.  In the
model III, no discrete symmetry is imposed and both up- and
down-type quarks may couple with either of the two Higgs doublets.
The Yukawa coupling in quark sector in this case is
\cite{atwood97}
\beq {\cal L}_Y = \eta_{ij}^U \bar{Q}_{i L}
\tilde{\phi_1} U_{j R}+\eta_{ij}^D \bar{Q}_{i L} \tilde{\phi_1}
D_{j R}+\xi_{ij}^U \bar{Q}_{i L} \tilde{\phi_2} U_{j R}+\xi_{ij}^D
\bar{Q}_{i L} \tilde{\phi_2} D_{j R}+  h.c ~,   \label{lagr}
\eeq
where $ \phi_1 $  and $\phi_2 $ are the two Higgs doublets and
$\eta_{i j}^{ U, D} $ and $ \xi_{i j}^{U, D} $ are the Yukawa
couplings. As described in \cite{atwood97}, in order to let $
\phi_1 $ correspond to generate fermion masses while $ \phi_2 $ to
introduce new interactions, one can choose the following
parameterization of the Higgs doublets
\beq
\phi_1 =\frac{1}{\sqrt{2}} \left [\left ( \begin{array}{c} 0 \\ v +H^0
\end{array} \right )+  \left (\begin{array}{c} \sqrt{2} \chi^+  \\
i \chi^0
\end{array} \right ) \right ]~,\ \ \ \ \ \
\phi_2=\frac{1}{\sqrt{2}} \left( \begin{array}{c} \sqrt{2} H^+
\\ H^1+i H^2
\end{array} \right ) ~,   \label{higgs}
\eeq
where $ v = (\sqrt{2} G_F)^{-1/2} =246 GeV $. After the
rotation that diagonalizes the mass  matrices of quark fields and
that of the Higgs doublets, we can get the Lagrangian that is
relevant for our following discussions. The main features of
model III are as follows \cite{atwood97}:
\begin{itemize}

\item
FCNC may exist at the tree level. The neutral and the charged
flavor changing couplings are related by:
\beq
\xi_{charge}^U=\xi_{neutral}^U  V_{CKM}~, \ \ \  \
\xi_{charge}^D=V_{CKM} \xi_{neutral}^D~,
\eeq
where $V_{CKM}$ is the ordinary mixing matrix between down-type
quarks\cite{ckm}, and
\beq
\xi_{neutral}^{U,D} = (V_L^{U, D})^{-1} \xi^{U,D} V_R^{U, D}.
\eeq

\item
Like the models I and II, there are also five Higgs bosons in model III:
 the light and heavy CP-even neutral Higgs boson $ h^0 $ and $\bar{H}^0$,  one CP-odd
neutral Higgs boson $ A^0 $ and a pair of charged Higgs bosons $ H^\pm $.
The transformation relation between $ ( H^0, H^1, H^2 ) $ in
Eq.(\ref{higgs}) and the mass eigenstates $( \bar{H}^0, h^0, A^0 )
$ can be found in \cite{atwood97}.

\end{itemize}

In the following of this paper, we will parameterize  $ \xi_{neutral}^{U,D} $
as Ref. \cite{atwood97}
\beq
(\xi_{neutral}^{U,D})_{i j}= \lambda_{ij}^{U, D} \frac{\sqrt{m_i m_j}}{v}~,
\eeq
and treat $ \lambda_{i j}^{U, D} $ as basic free  parameters.

\subsection{Experimental constraints}

There is a considerable interest in the constraint of the
parameter space of the 2HDM, especially in model III, since the
FCNC may appear at the tree level. Compared with the SM, the
additional free parameters of model III are the masses of the
additional Higgs bosons and the coupling constants $ \lambda_{ij}
$. In this subsection, we summarize the main constraints on these
parameters from direct searches at LEP experiments\cite{pdg2000},
and from the measurements of $F^0-\bar{F}^0 $ mixing with $F^0=K^0,
B_d^0$.

Let's firstly turn to mass constraints. The LEP Working Group for
Higgs boson searches \cite{lephiggs} has recently reported
excess of events that might be due to the production of a
neutral Higgs boson weighing about $115$ GeV, and placed a lower
mass limit for the SM Higgs boson $M_{H^0} >113.5 $ GeV at $95\%
C.L.$. For the $h^0$ and $A^0$ Higgs bosons of the MSSM, the new
$95\% C.L. $ limit\cite{ep0055} is $M_{h^0} > 88.3$ GeV and
$M_{A^0}> 88.4$ GeV respectively. But one should note that these
constraints on the masses of $h^0$ and $A^0$ are not applicable in
the 2HDM's because the coupling of $h^0$ to $Z^0 Z^0$ and to $A^0
Z^0$ go like $\sin \alpha$ and $\cos \alpha$( $\alpha$ is a free
parameter of the model), respectively. For a very small mixing
angle $\alpha$ and with $A^0$ ($h^0$) sufficiently heavy no lower
limit on $M_{h^0}$ ($M_{A^0}$) can be set from the LEP data. For
the charged Higgs bosons in the 2HDM's, the $95\% C.L. $ limit is
$M_{H^\pm} > 78.6$ GeV\cite{ep0055}.

Indirect constraints on Higgs masses come from $R_b$ and $\rho$,
which have been measured at LEP. According to studies in
Ref.\cite{cck99}, if one requires $R_b^{theory} $ to be within $ 1
\sigma $ deviation from $ R_b^{exp} $ and $\rho $ within $ 2
\sigma $ deviation from $ \rho^{exp} $, the preferred range of
Higgs boson masses in model III are $ 80 GeV < M_{h^0} \simeq
M_{A^0} < 120 GeV $ and $ M_{H^\pm} \approx (180- 220) GeV
$\cite{cck99}. On the other hand, if one allows $R_b^{theory} $ to
vary within $ 2 \sigma $ errors of $ R_b^{exp} $, a charged Higgs
mass of a few hundreds GeV is feasible. In addition, unitary
condition requires these masses to be less than $ 1 TeV
$\cite{unit}.

As for the couplings $ \lambda_{i j}^{U, D} $, a lot of processes
have to be analyzed to give reasonable constraints, as have been
done in Refs.\cite{atwood97,cck99}. In Table \ref{table3}, we list
current constraints on these couplings along with the processes
from which constraints have been or will be placed. It should be
noted that despite current constraints on flavor changing
couplings $\lambda_{b s}^{D} $ and $ \lambda_{t c}^{ U} $ are
rather loose, some much more stringent constraints will be imposed
in the future experiments.

For simplicity, we set (a)
all the FC couplings $ \lambda_{i j}^{U, D} (i \neq j) =0 $;
(b) $ \lambda_{i i}^{U} = \lambda_1 $  for $i = u,c ,t, $ and $
\lambda_{i i}^{D} = \lambda_2 $ for $ i=d, s, b$;
and (c) $ \lambda_{1,2} $ are real numbers.
The advantage of
such a setting is the new contributions to Wilson coefficients
discussed below come only from the charged Higgs penguin diagrams
and the neutral Higgs bosons are irrelevant to our discussion. The
model III therefore differs from the model II only in the
couplings of the charged Higgs boson to fermions. The Feynman
rules for  the $\bar{U_i}H^+D_j$ and $\gamma H^+ H^-$ vertex are
\beq
\bar{U_i}H^+D_j: && \ \ \ \frac{- i}{2} \left [ V_{CKM}\cdot
\xi^D_{ij}(1+\gamma_5) -\xi^U_{ij}\cdot V_{CKM}(1-\gamma_5) \right
]~, \\ \gamma H^+ H^-: && \ \ \  i e (p_2-p_1)^{\mu}~,
\eeq
For more details about the couplings between the Higgs bosons and
quarks or gauge bosons, one can see Ref.\cite{atwood97}.

Another very important source of constraint is the process $ B \to
X_s \gamma $. In the SM, the theoretical prediction of ${\cal B} (
B \to X_s \gamma ) $ is $(2.8 \pm 0.8 ) \times 10^{-4} $ at LO
approximation \cite{bsg}, and $ (3.28 \pm 0.33 ) \times 10^{-4} $
at NLO approximation \cite{nbsg}. The $95 \% C.L.$ limit from CLEO measurement
\cite{cleo99} is $ 2 \times
10^{-4} < {\cal B}  ( B \to X_s \gamma ) < 4.5 \times 10^{-4} $.
One can see that the theory and the data are in
remarkable agreement, which implies that only a narrow parameter space
of new physics can survive. Since the aim of this paper is to estimate
the new physics effect on the process $ B \to X_s \gamma \gamma$,
we will vary  the input parameters of 2HDM  in a rather large range
allowed by
\beq
\frac{2.0 \times 10^{-4}} {2.8 \times 10^{-4}}
<\frac{{\cal B} _{III}^{LO} (B \to X_s \gamma )} {{\cal B} _{SM}^{LO} (B \to
X_s \gamma )} \left |_{\mu =m_b}  < \frac{4.5 \times 10^{-4}} {2.8
\times 10^{-4}} \right. ~. \label{constr}
\eeq

\subsection{Calculation of $B \to X_s \gamma \gamma$}

In model III, the operator bases presented in Eq.(\ref{operator})
are insufficient for the process $ B \to X_s \gamma \gamma $ and
we need following additional operators  when all the FC couplings
are zero \cite{ai98}: \footnote{Here we neglect the operators
$O_{9, 10} $ and $O_{9, 10}^{\prime} $ introduced in \cite{cm94}
because in our case, the corresponding Wilson coefficients keep zero at any
scale.}
\beq
O_1^{\ \prime} &=&(\bar{s}_\alpha \gamma^{\mu} R
c_\beta)(\bar{c}_\beta \gamma_{\mu} R b_\alpha)~, \nonumber \\
 O_2^{\ \prime}&=&(\bar{s}_\alpha
\gamma^{\mu} R c_\alpha)(\bar{c}_\beta \gamma_{\mu} R b_\beta)~, \nonumber \\
 O_{3, 5}^{\ \prime}&=&(\bar{s}_\alpha
\gamma^{\mu} R b_\alpha) \sum_{q=u \cdots b}(\bar{q}_\beta
\gamma_{\mu} (R, L) q_\beta)~, \nonumber \\
 O_{4, 6}^{\ \prime}&=&(\bar{s}_\alpha
\gamma^{\mu} R b_\beta) \sum_{q=u \cdots b}(\bar{q}_\beta
\gamma_{\mu} (R, L) q_\alpha)~, \nonumber \\
O_{7}^{\prime}&=&\frac{e}{16 \pi^2} \bar{s}_{\alpha} \sigma^{\mu \nu} (
m_b L + m_s R) b_{\alpha} F_{ \mu \nu}~,   \nonumber  \\
O_{8}^{\prime}&=&\frac{g_s}{16 \pi^2} \bar{s}_{\alpha} \sigma^{\mu \nu} (
m_b L + m_s R)\lambda_{\alpha \beta}^{a} b_{\beta} G_{ \mu \nu}^a~.
\eeq

\begin{figure}
\begin{center}
\begin{picture}(300,200)(0,0)
\put(-60,-140) {\epsfxsize140mm\epsfbox{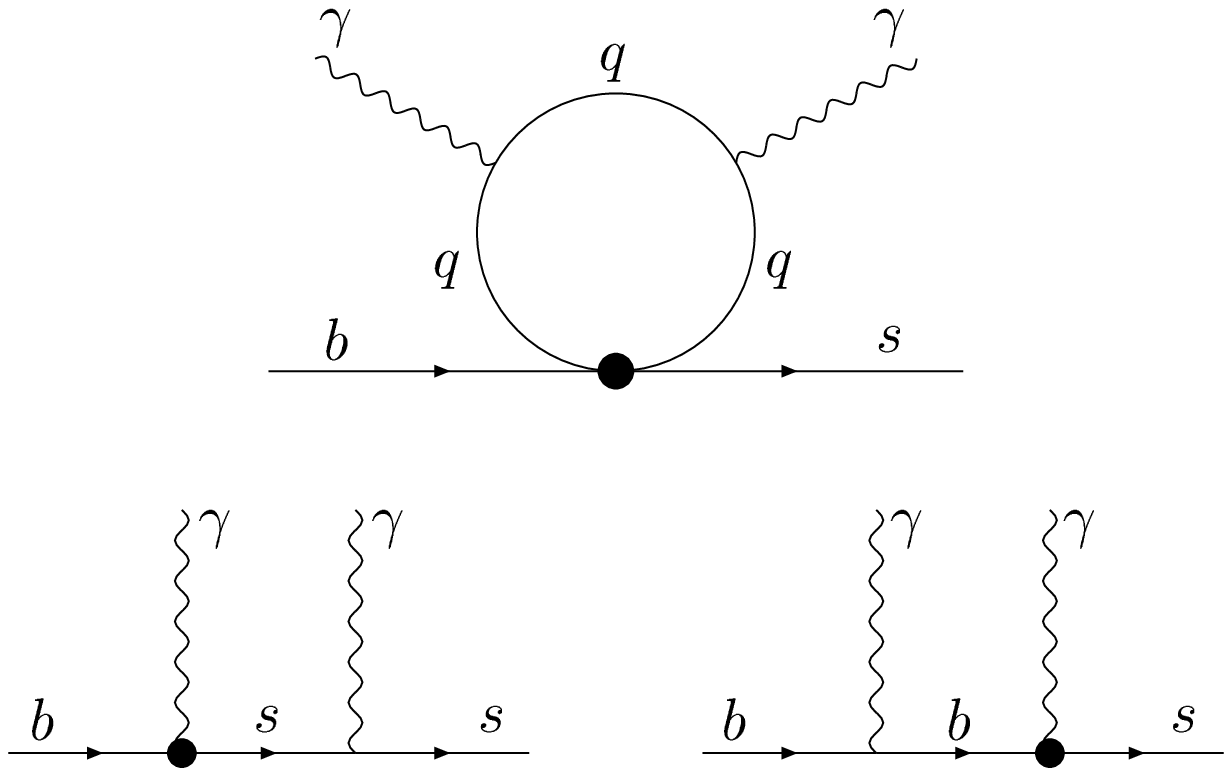}}
\end{picture}
\caption{Examples of Feynman diagrams that
contribute to the matrix elements
$ < s \gamma \gamma | H_{eff} | b> $. The 1PI diagrams illustrate possible insertion of
$ O_1 \sim O_6 $, while the 1PR diagrams represent the insertion of $O_7 $.
\label{fig1}}

\begin{picture}(300,300)(0,0)
\vspace{80pt}
\put(-60,0){\epsfxsize140mm\epsfbox{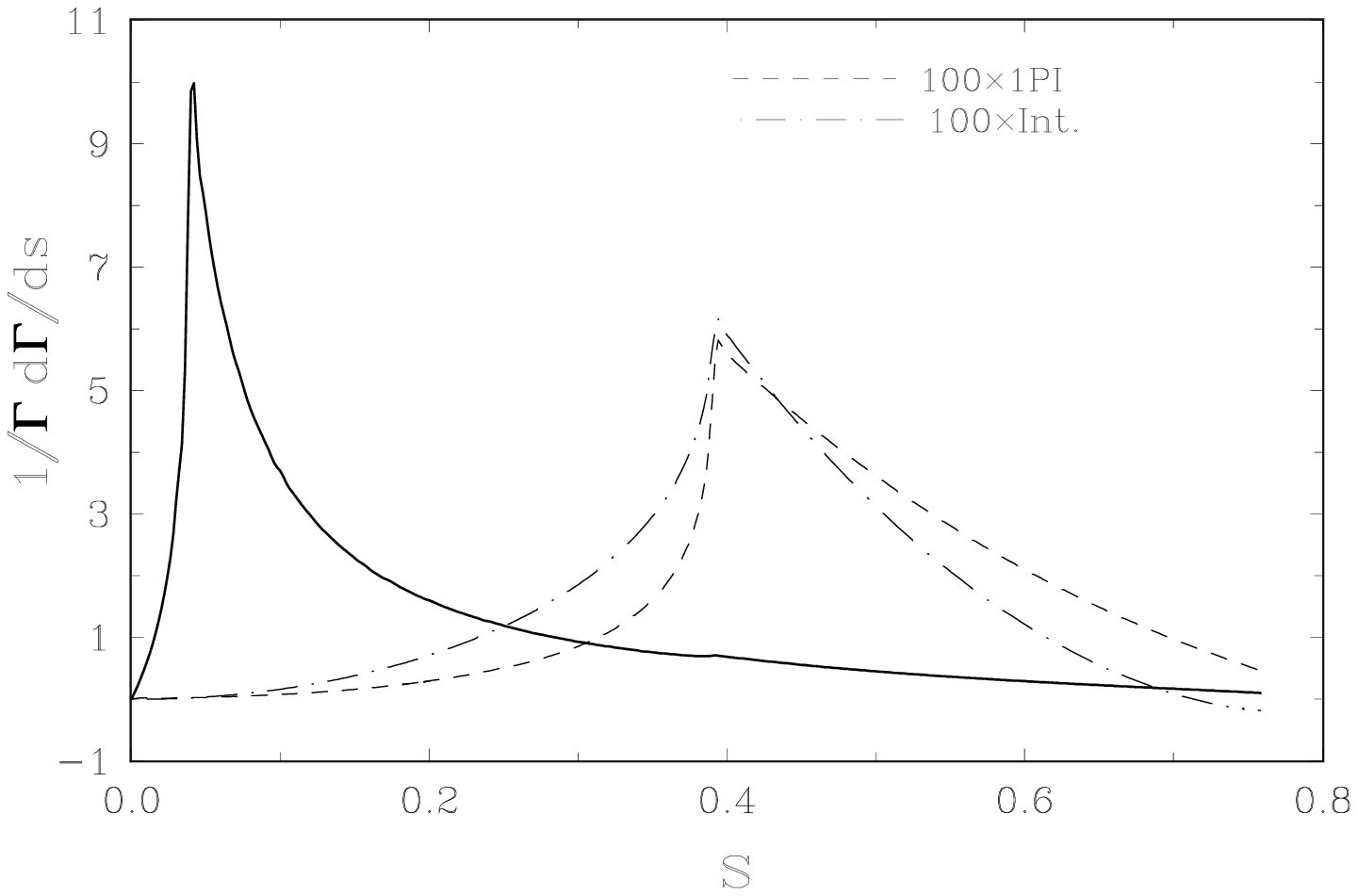}}
\end{picture}
\caption{The normalized distribution $1/\Gamma d\Gamma/d s $ versus $ s $ in
the SM (solid line). The contribution from 1PI diagrams ( dashed line) and that
from the interference between 1PI and 1PR diagrams (dot dashed
line) are also shown. In plotting this diagram,
we set $ \mu = m_b $. The mean value of $ s $ is 0.18 for the cuts as
given in the text. \label{fig2}}

\begin{picture}(300,250)(0,0)
\put(-60,0){\epsfxsize140mm\epsfbox{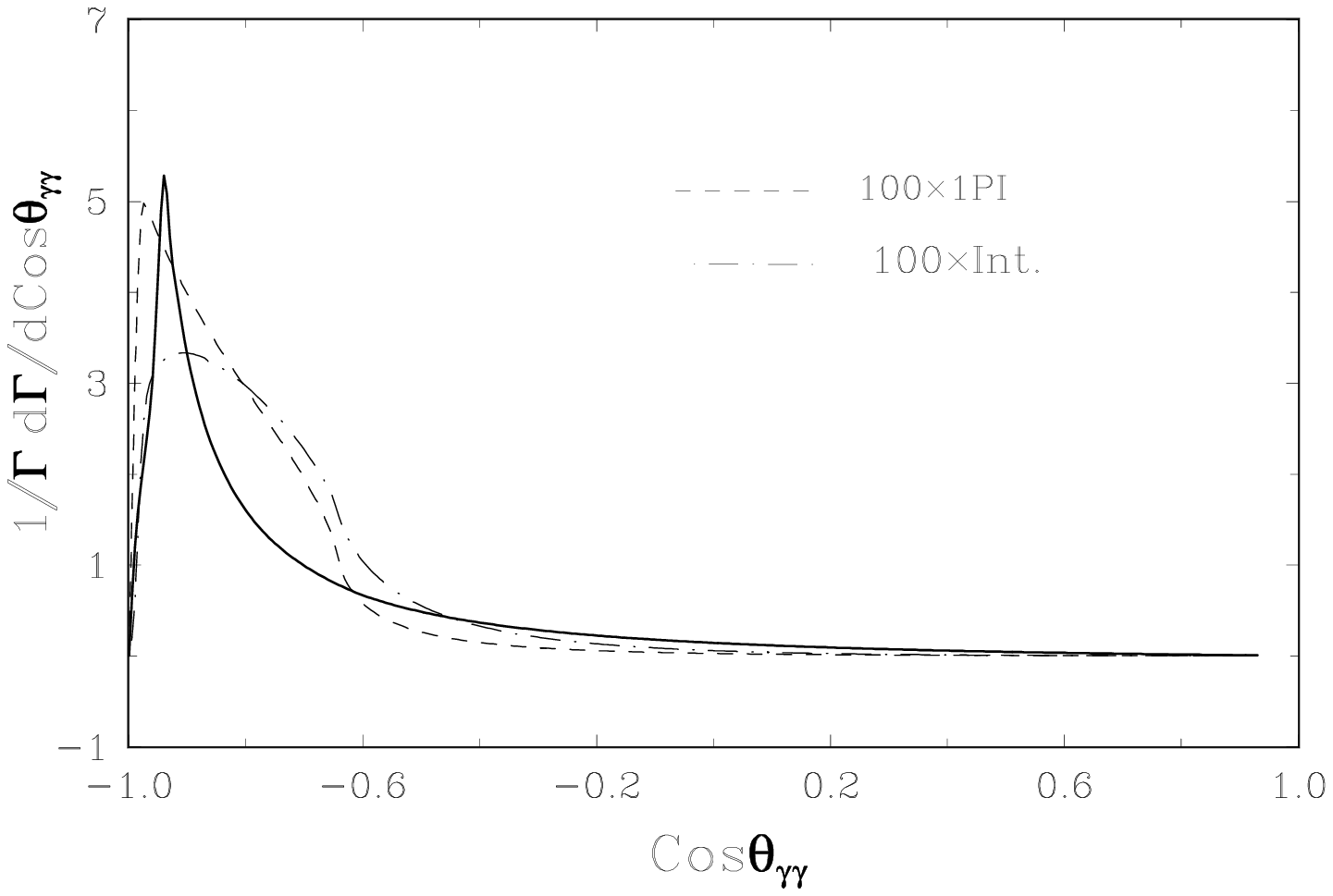}}
\end{picture}
\caption{The normalized distribution $1/\Gamma
d\Gamma/d(\cos\theta_{\gamma \gamma}) $ versus $
\cos\theta_{\gamma \gamma} $ in the SM (solid curve) for fixed
$\mu = m_b$. The contribution from 1PI diagrams (dashed curve) and
the interference between 1PI and 1PR diagrams (dot-dashed curve)
are also plotted. The mean value of $ \cos \theta_{\gamma \gamma}
$ is -0.70 for the cuts as given in the text. \label{fig3}}

\begin{picture}(300,250)(0,0)
\vspace{80pt}
\put(-60,0) {\epsfxsize140mm\epsfbox{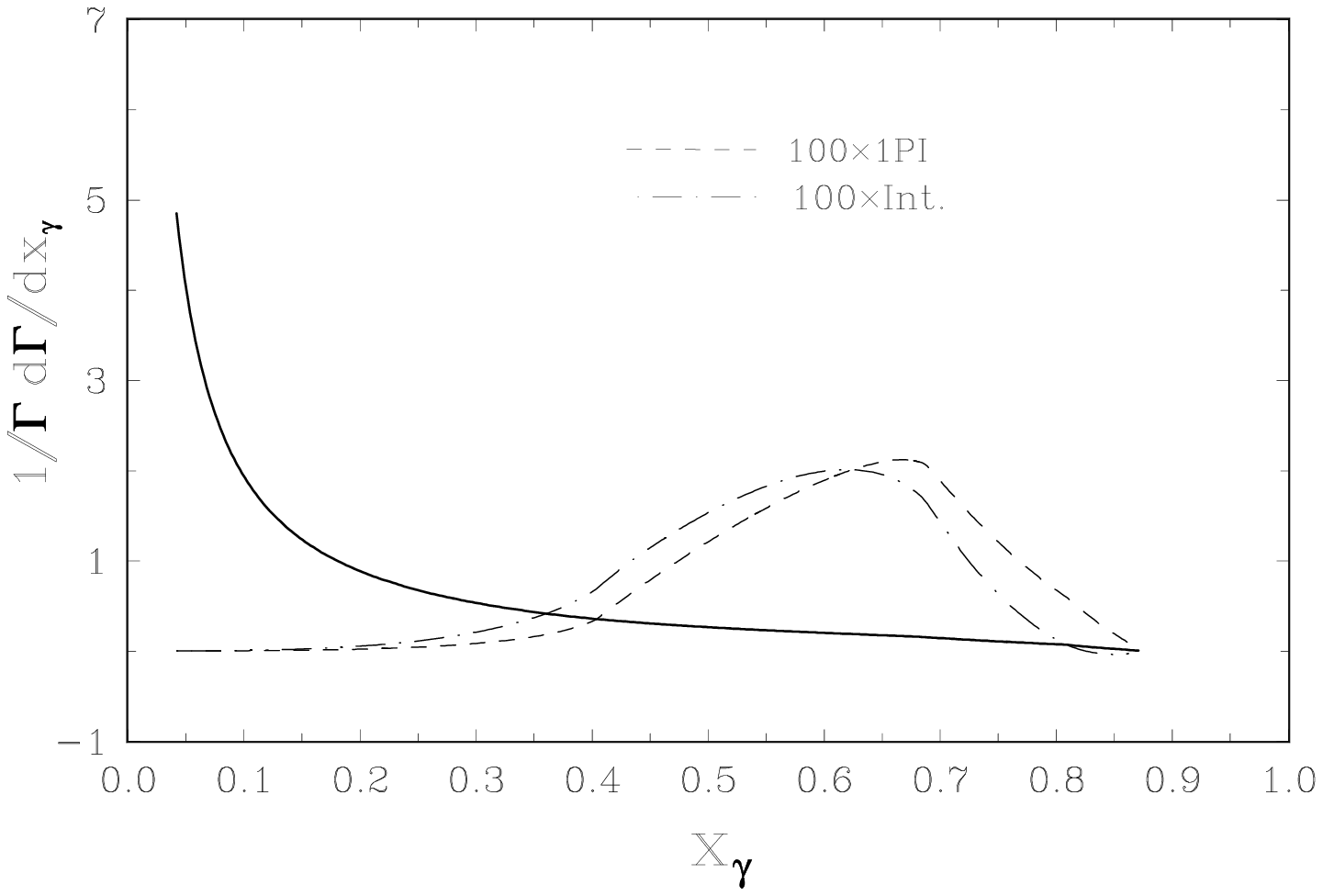}}
\end{picture}
\caption{The spectra of lower energy photons (solid curve) in the
SM. The contribution from 1PI diagrams (dashed curve) and that from the
interference between 1PI and 1PR diagrams (dot-dashed curve ) are
also shown. In plotting this diagram, we set $ \mu = m_b $, $ X_{\gamma}=2 E_{\gamma}
/ m_b $ and normalize the spectra to the total QCD corrected rate. \label{fig4}}

\begin{picture}(300,270)(0,0)
\put(-60,0) {\epsfxsize140mm\epsfbox{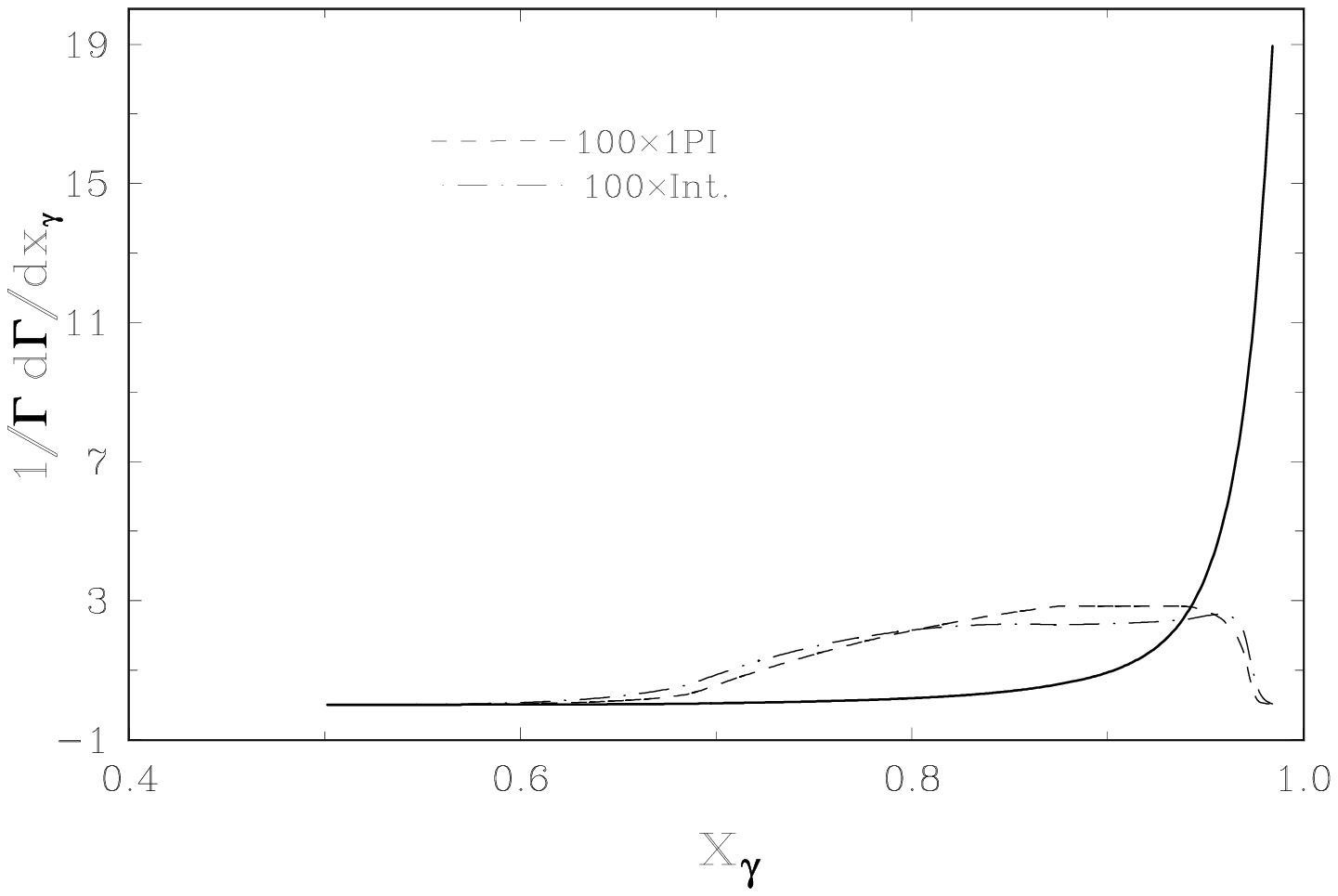}}
\end{picture}
\caption{Same as Fig.4, but for the spectrum of high energy photon.  \label{fig5}}

\begin{picture}(300,220)(0,0)
\put(-60,-160) {\epsfxsize150mm\epsfbox{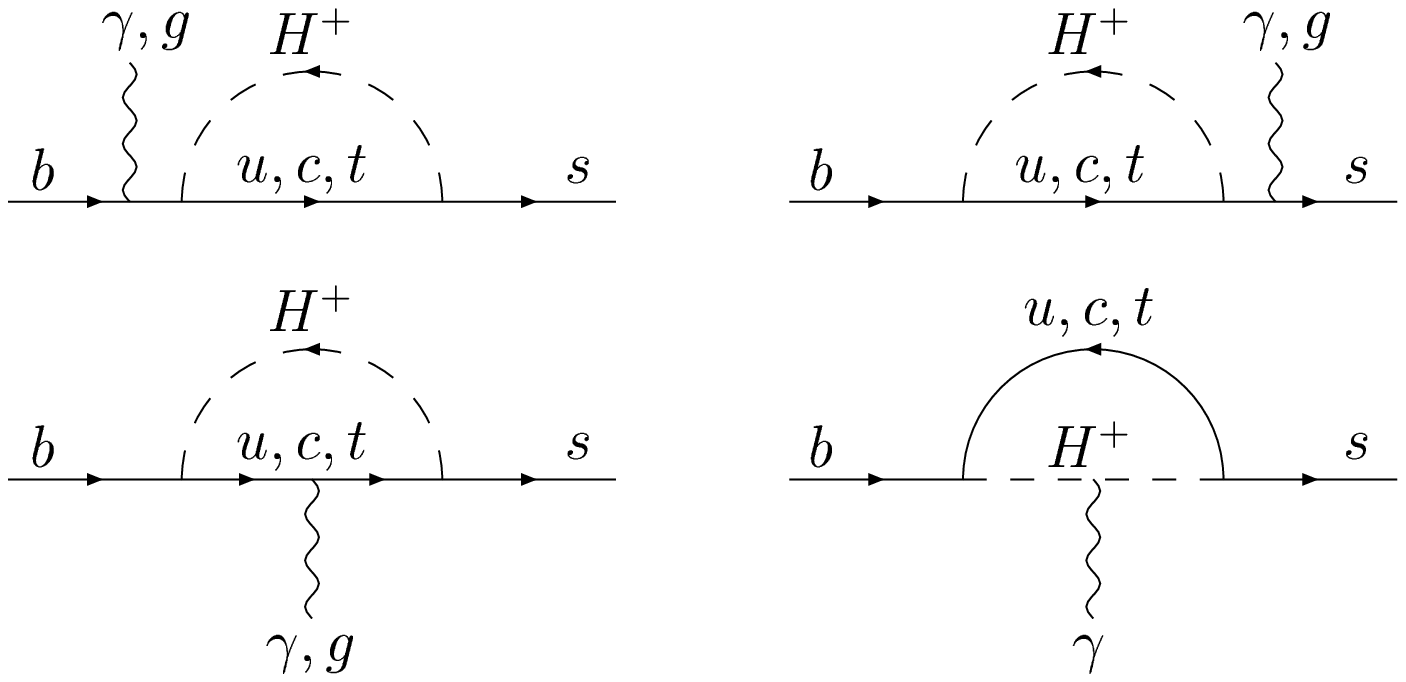}}
\end{picture}
\caption{ The Feynman diagrams relevant to the decays $ b \to s
\gamma $ and $ b \to s g $ in model III.  From these diagrams,
$C_i^{H} $ and $ C_i^{\prime \  H} $ can be extracted.  The
internal quarks are the upper type $ u $, $ c $ and $ t $ quarks.
\label{fig6}}

\begin{picture}(300,250)(0,0)
\put(-60,0){\epsfxsize140mm\epsfbox{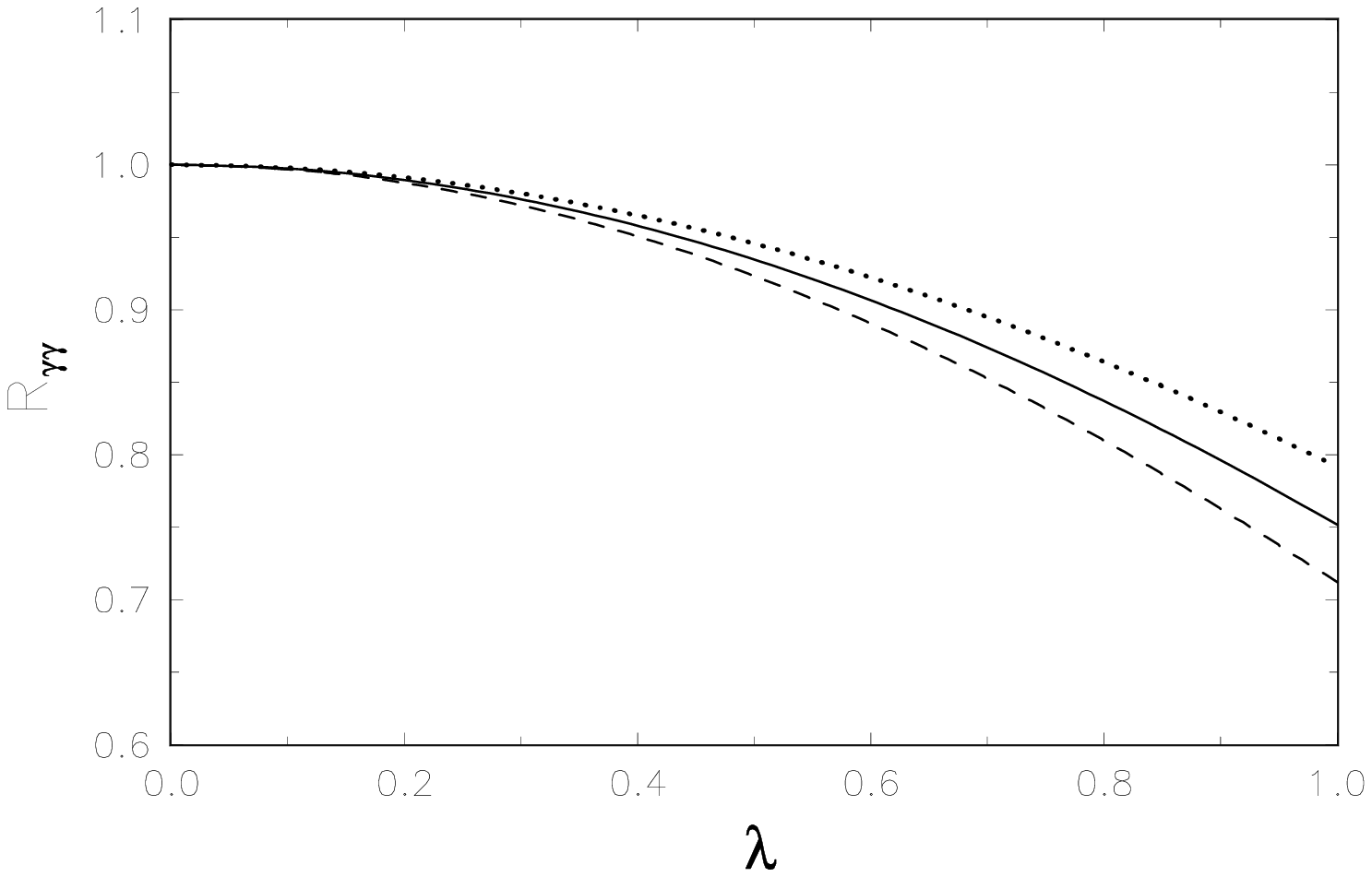}}
\end{picture}
 \caption{$R_{\gamma \gamma} $ as a function of the coupling $ \lambda $ in
the Case-1 of Scenario I
 for fixed $m_{H^{\pm}}=300 GeV $.  The dashed, solid and dotted curve correspond to
 $ \mu=2 m_b $, $m_b $ and $ m_b/2 $, respectively.
  In determining the range of $ \lambda $, we have required Eq.(\ref{constr}) to be
  satisfied and $C_7$ to be negative. \label{fig7}}

\begin{picture}(300,300)(0,0)
\put(-60,0){\epsfxsize140mm\epsfbox{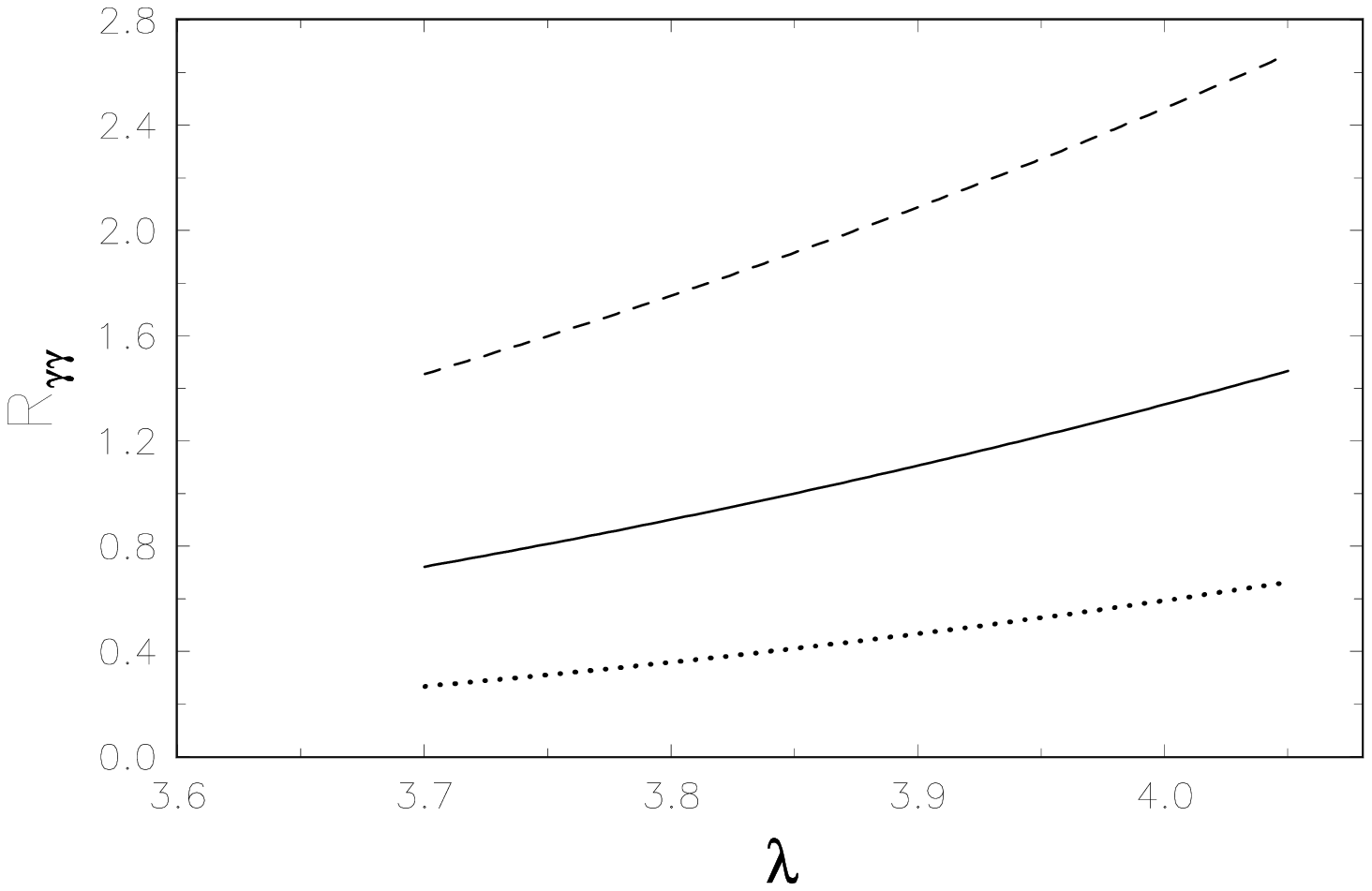}}
\end{picture}
 \caption{Same as Fig.7, but for a positive $C_7$. \label{fig8}}

\begin{picture}(300,250)(0,0)
\put(-60,0){\epsfxsize140mm\epsfbox{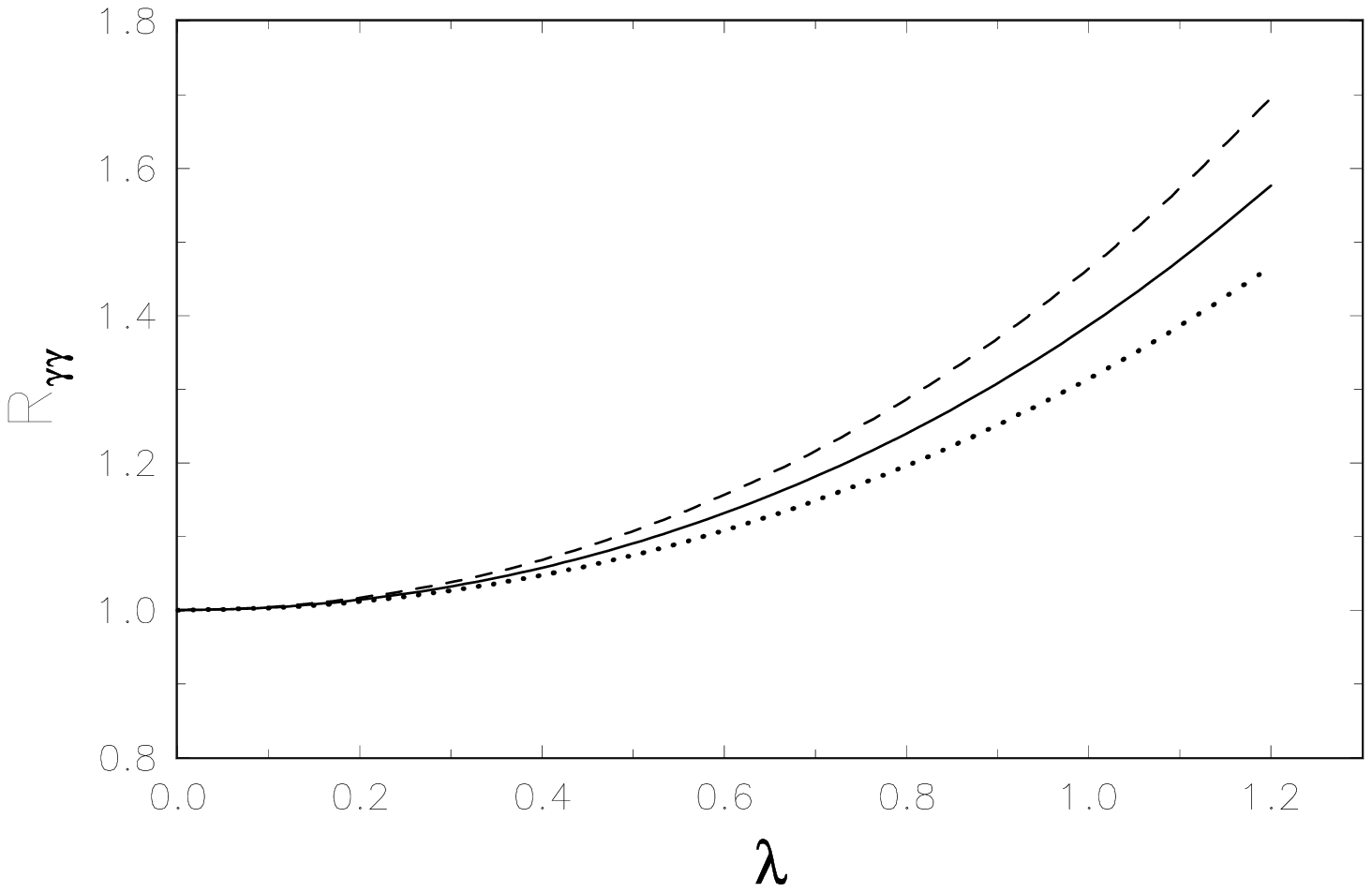}}
\end{picture}
 \caption{ Same as Fig.7, but for the Case-2 of Scenario I. In this case,
  $C_7$ keeps negative for all the values of $\lambda$. \label{fig9}}

\begin{picture}(300,300)(0,0)
\put(-60,0){\epsfxsize140mm\epsfbox{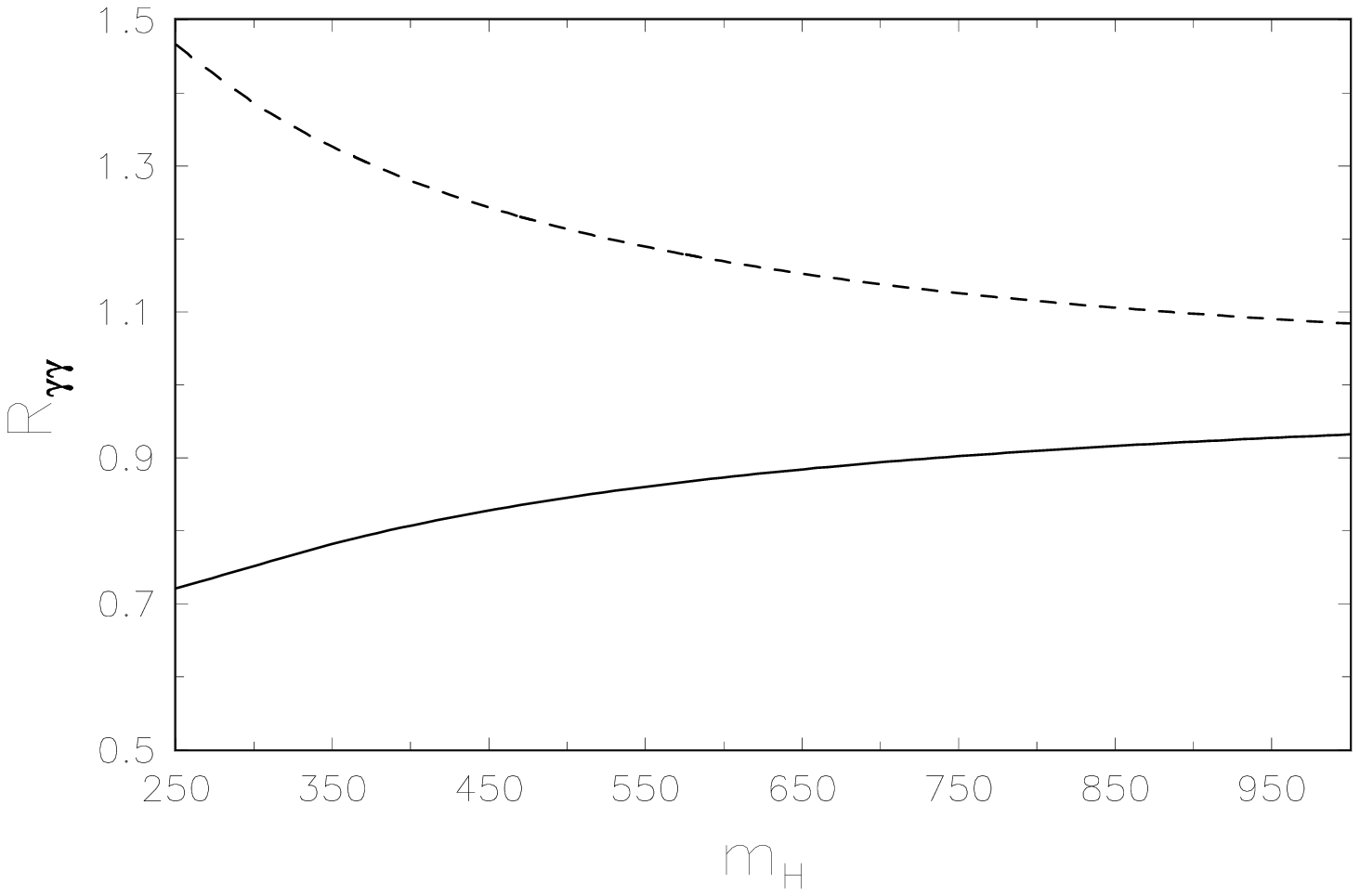}}
\end{picture}
 \caption{The dependence of $R_{\gamma \gamma} $ on
 the charged Higgs mass $ m_{H^{\pm}} $
for Case-1 (solid curve) and Case-2 (dashed curve)  in scenario I.
In either case, $C_7 $ is negative and we set $\lambda =1 $
and $ \mu=m_b $. \label{fig10}}

\begin{picture}(300,250)(0,0)
\put(-60,0){\epsfxsize140mm\epsfbox{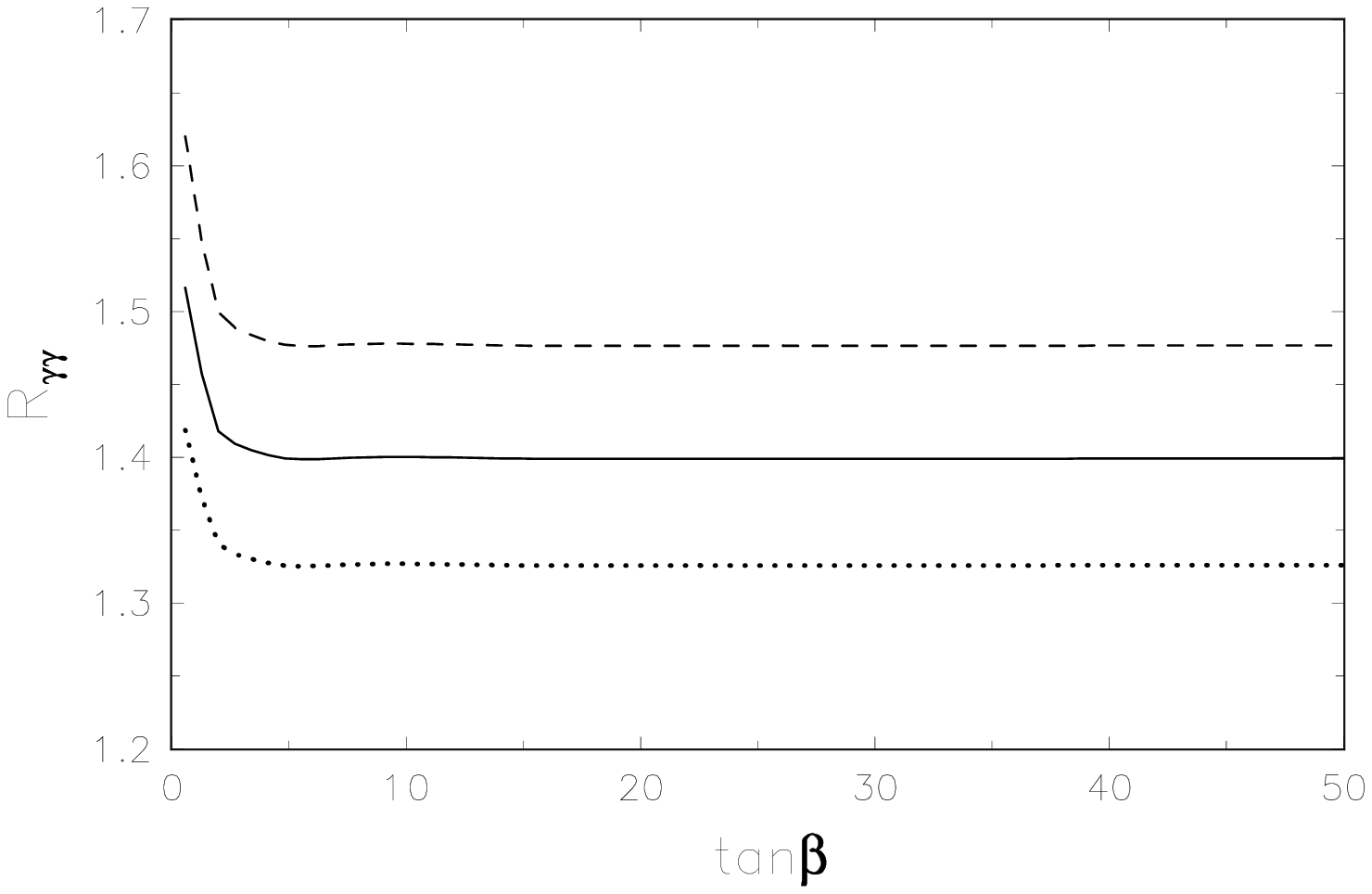}}
\end{picture}
\caption{$\tan{\beta}$ and $\mu$ dependences of the ratio $R_{\gamma \gamma} $ in
model II for fixed $\mhp =500$ GeV.  The dashed, solid and dotted curves correspond
to $ \mu=2 m_b $, $m_b $,  and $ m_b/2 $, respectively.
In numerical calculation, we have required the constraints as given in
Eq.(\ref{constr}) to be satisfied.  \label{fig11}}

\begin{picture}(300,250)(0,0)
\vspace*{-50pt}
\put(-60,0){\epsfxsize140mm\epsfbox{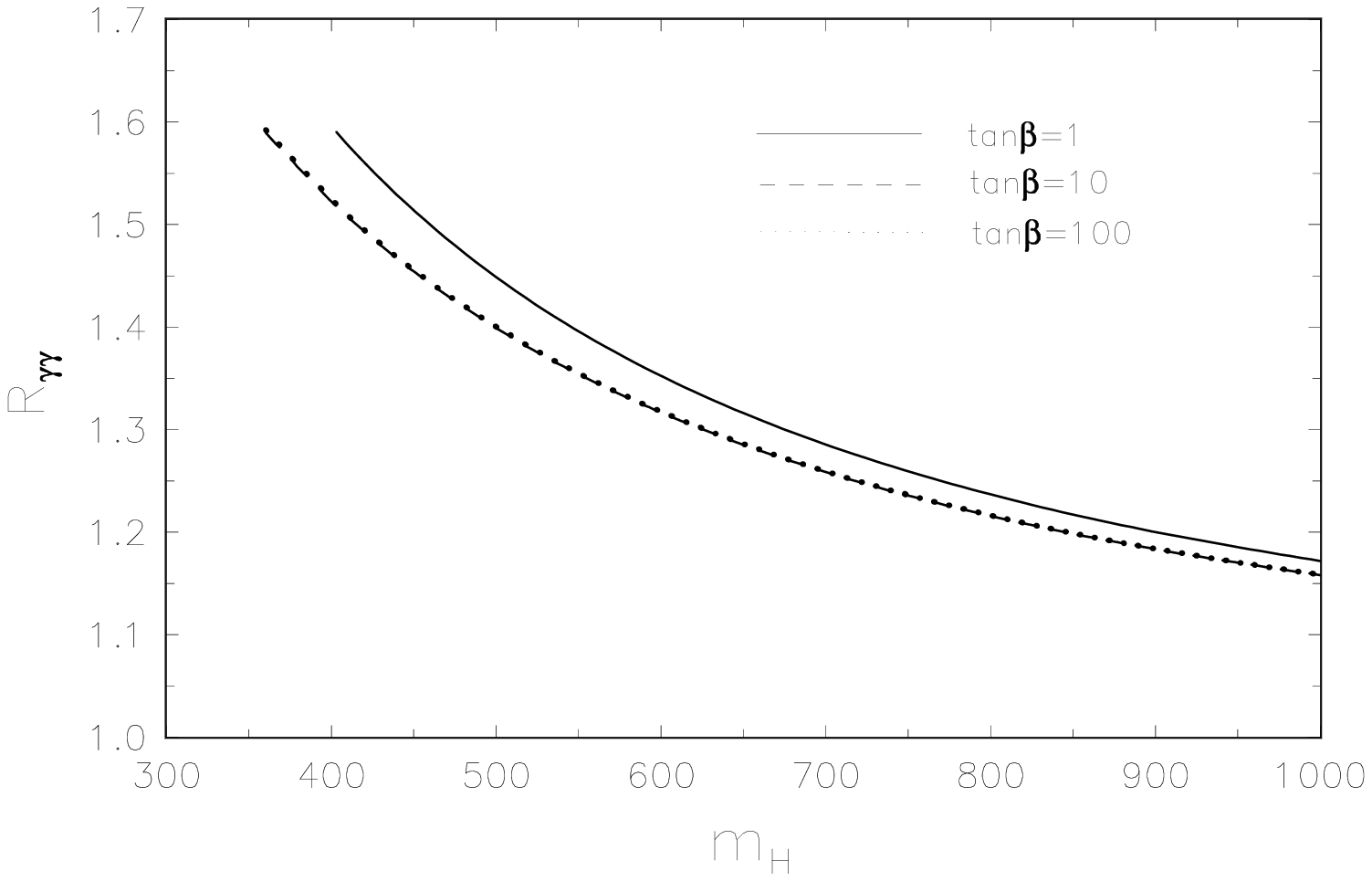}}
\end{picture}
\caption{$\mhp$ and $\tan{\beta}$ dependences of $R_{\gamma \gamma} $ in model II
for fixed $\mu=m_b$.  The soild, dashed and dotted curves correspond to
$ \tan{\beta}=1, 10, 100$, respectively. The last two curves can not be
separated clearly.
In numerical calculation, we have required the constraints as given in
Eq.(\ref{constr}) to be satisfied.  \label{fig12}}

\begin{picture}(300,280)(0,0)
\vspace*{-100pt}
\put(-60,0){\epsfxsize140mm\epsfbox{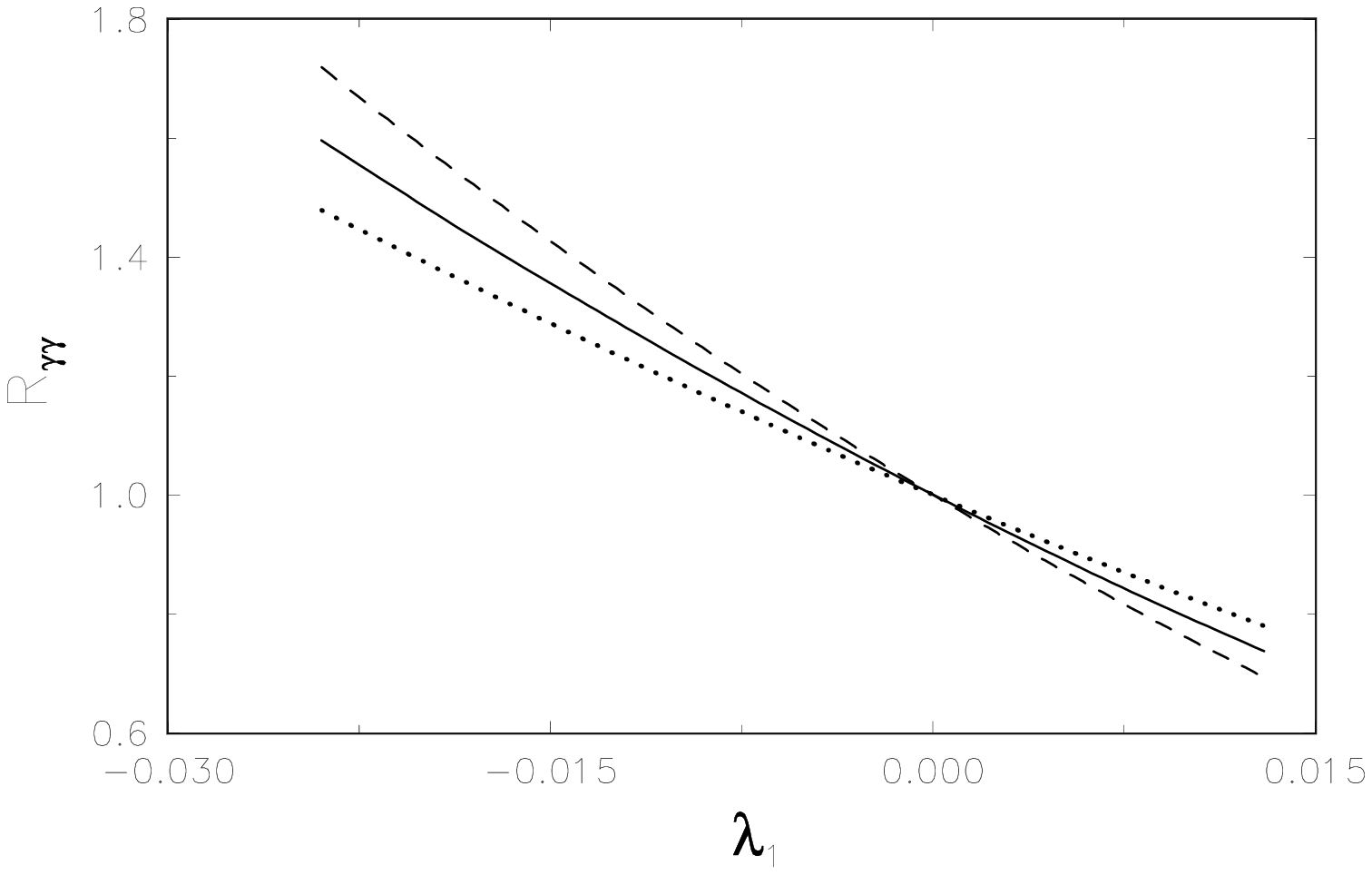}}
\end{picture}
 \caption{The dependence of $R_{\gamma \gamma} $ on  $ \lambda_1 $ in scenario III for fixed $ \lambda_2 =50 $
 and $ m_{H^{\pm}}=200 GeV $. The dashed, solid and dotted curves corresponds to $ \mu=2 m_b$,  $ m_b $and $m_b/2 $
 respectively. In determining the range of $ \lambda_1 $, we required Eq.(\ref{constr}) to be
 satisfied and $C_7 $ to be negative. \label{fig13} }

\begin{picture}(300,250)(0,0)
\put(-60,0){\epsfxsize140mm\epsfbox{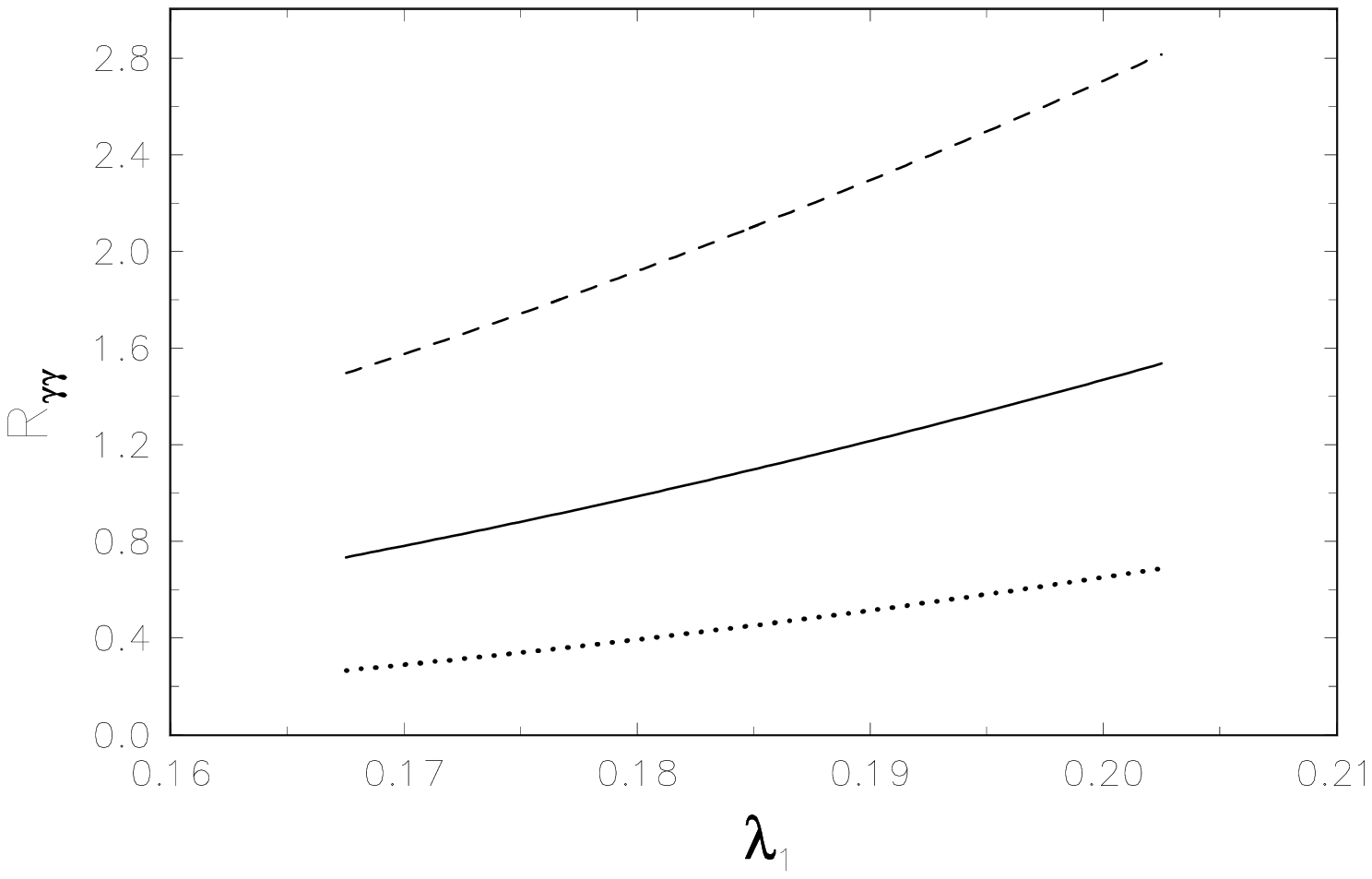}}
\end{picture}
\caption{Same as Fig.\ref{fig13},
but for a positive $C_7$. \label{fig14}}

\begin{picture}(300,250)(0,0)
\vspace{-50pt}
\put(-60,0){\epsfxsize140mm\epsfbox{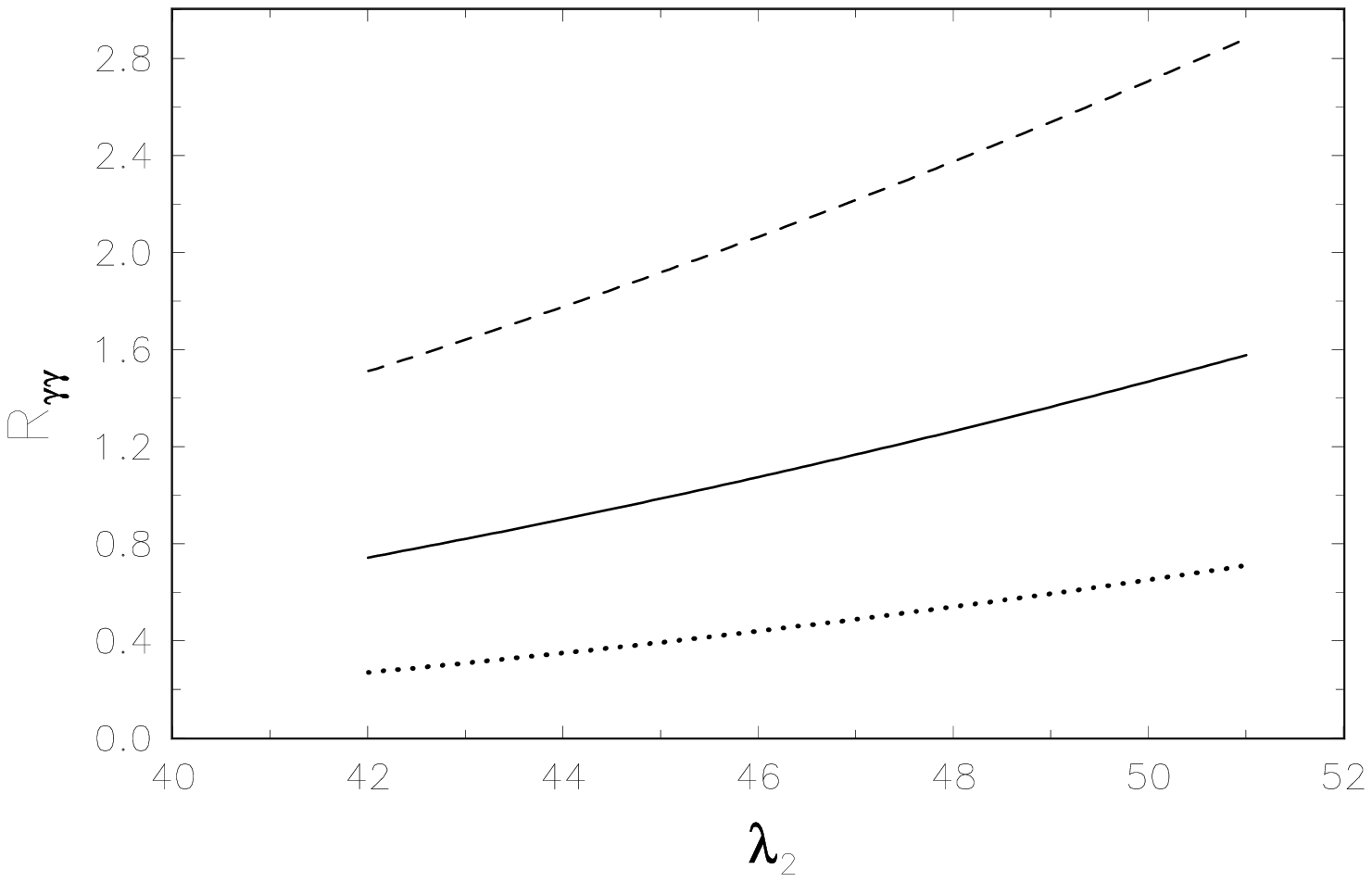}}
\end{picture}
 \caption{The relationship between $R_{\gamma \gamma} $ and  $ \lambda_2 $
 in scenario III  for fixed $ \lambda_1 =0.2 $
 and $ m_{H^{\pm}}=200 GeV $. The dashed, solid and dotted curves corresponds
 to $ \mu=2 m_b$, $ m_b $and $m_b/2 $ respectively.
 In determining the range of $ \lambda_2 $, we have required Eq.(\ref{constr}) to
 be satisfied and $\lambda_2 \gg  \lambda_1  $. \label{fig15} }

\begin{picture}(300,250)(0,0)
\put(-60,0){\epsfxsize140mm\epsfbox{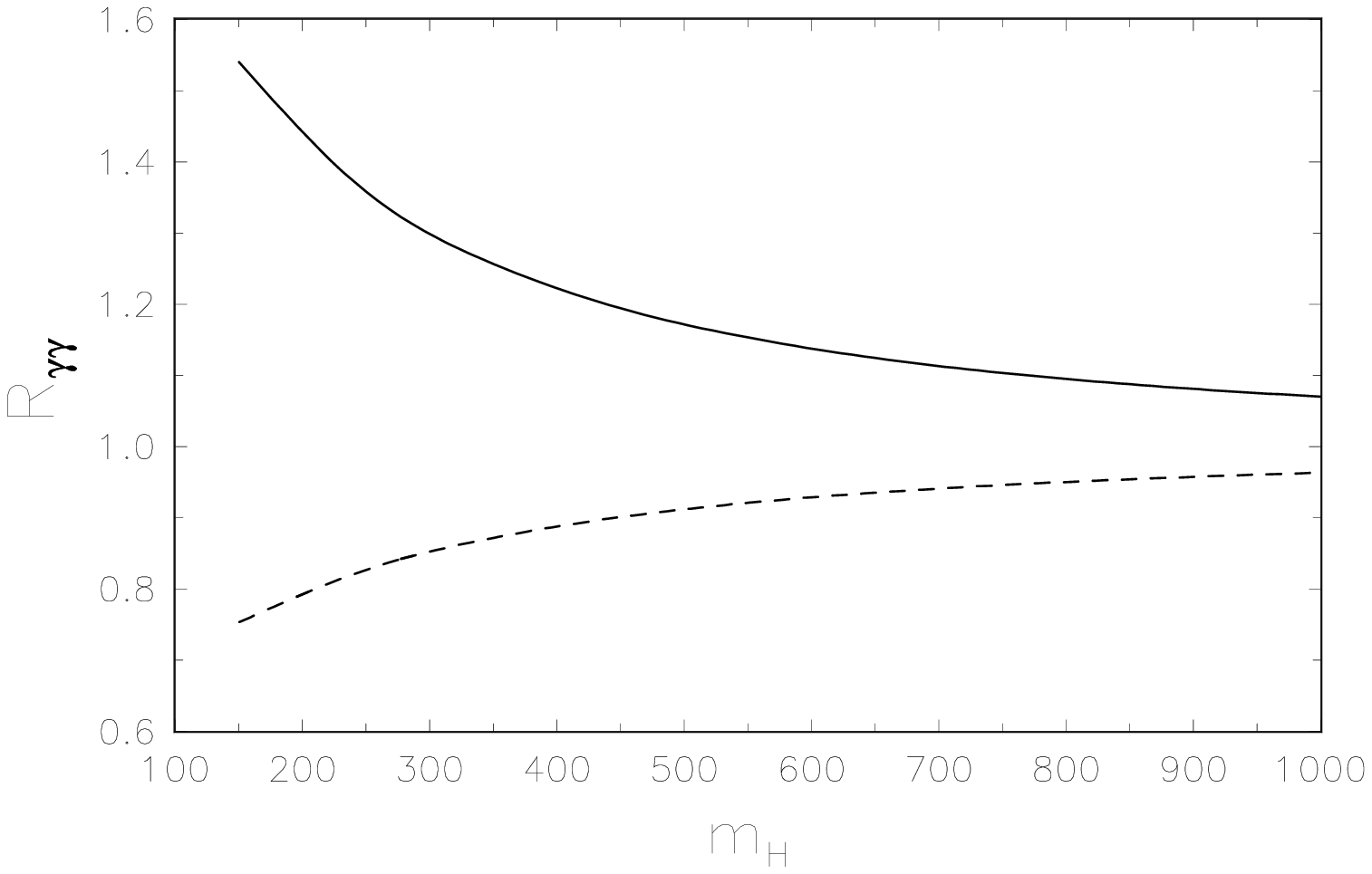}}
\end{picture}
 \caption{$R_{\gamma \gamma} $ as a function of
 the charged Higgs mass $ m_{H^{\pm}} $ for $ \lambda_2=50 $ and $ \mu = m_b $ .
 Solid and dashed curve correspond to $ \lambda_1=-0.018$
 and $0.010 $, respectively. When choosing the value of $ \lambda_1 $,
 we keep $ C_7 $ to be negative and require $ |C_7 |
>|C_7^{SM}| $ (solid curve) and $ |C_7 |< |C_7^{SM}| $ (dashed curve),
respectively.  \label{fig16}}

\begin{picture}(300,300)(0,0)
\put(-60,0){\epsfxsize140mm\epsfbox{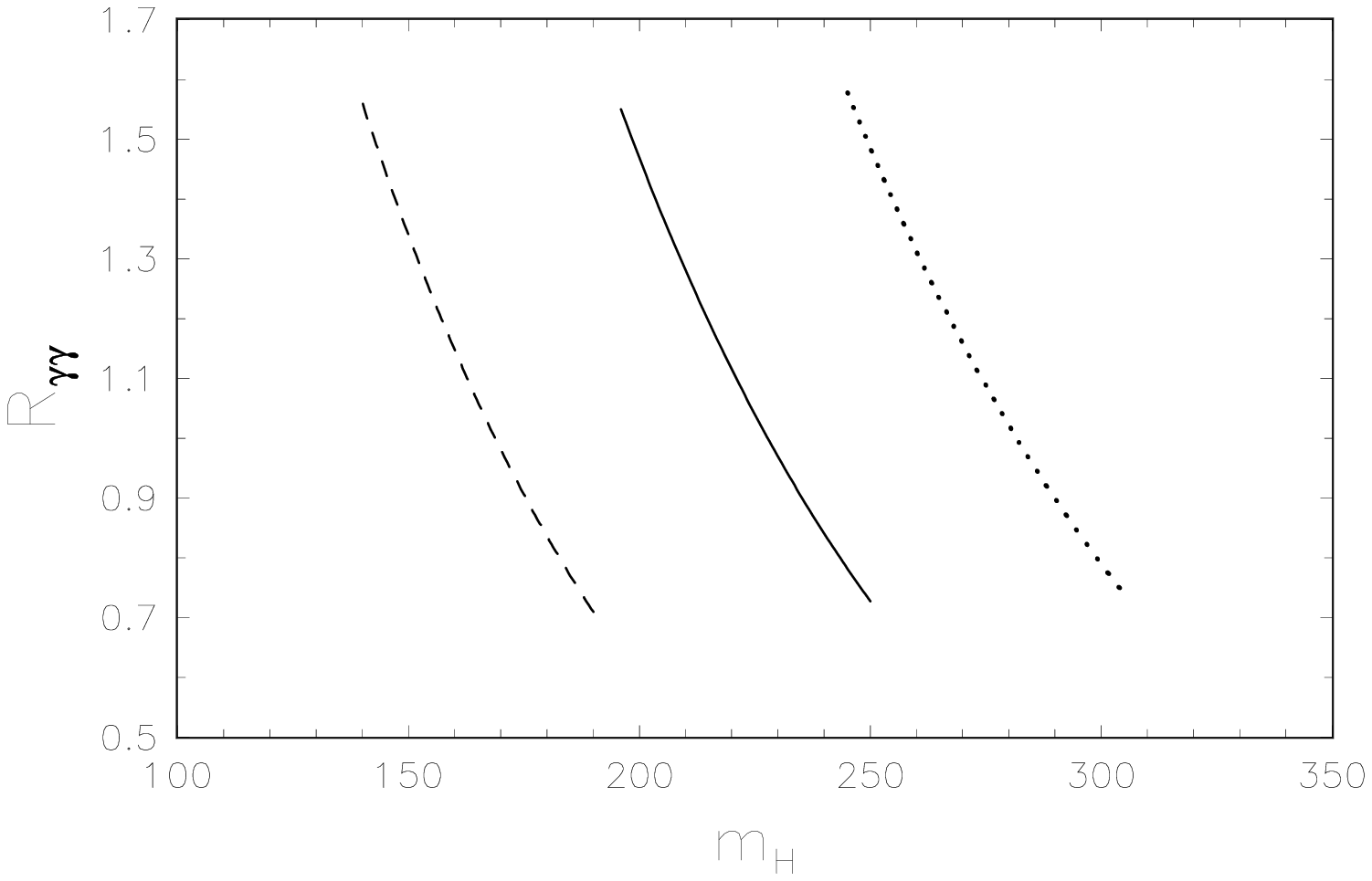}}
\end{picture}
 \caption{$ m_{H^{\pm}} $ dependence of $R_{\gamma \gamma} $ for
 fixed $ \lambda_1=0.2 $,  $ \mu=m_b $ and $\lambda_2=40$ (dashed curve), $50$ (solid
curve), $60$ (dotted curve).  To determin the range of
$m_{H^{\pm}} $, we require Eq.(\ref{constr})  to be satisfied and
$C_7 > 0 $. Numerical
 results also show  that for a large $m_{H^{\pm}} $, $C_7 $ is driven to be negative.
\label{fig17} }

\end{center}

\end{figure}

For the evaluation of Wilson coefficients, we need their initial
values with standard matching computations. Denoting the Wilson
coefficients in the SM with $ C_i^{SM}(m_W) $ and those from the
additional charged Higgs contribution (see Fig.\ref{fig6}) with  $
C_i^{H}(m_W) $, we have the initial values of the Wilson
coefficients for the first set of operators [ Eq. (\ref{operator}) ]
\beq
C_i^{III}(m_W)& = & C_i^{SM} (m_W) + C_i^{H}(m_W)~,
\eeq
where
\beq
C_{1 \cdots 6}^{H} (m_W)&= &0~, \non
C_{7}^{H}(m_W)&= &\frac{1}{2} \left [ \lambda_1^2 f_1(y) +\lambda_1
\lambda_2 f_2 (y) +\frac{m_s^2}{m_t^2} \lambda_2^2 f_1 (y) \right]~,
\non
C_{8}^{H} (m_W)&= &\frac{1}{2} \left [ \lambda_1^2 g_1 (y)+\lambda_1
\lambda_2 g_2 (y) +\frac{m_s^2}{m_t^2} \lambda_2^2 g_1 (y)  \right ]~,
\label{exprc7}
\eeq
and the explicit expressions of $
C_i^{SM}(m_W) $ can be found in the literature \cite{bbl96}. For
the primed Wilson coefficients we have
\beq
C_i^{\prime III} (m_W)=C_i^{\prime H} (m_W)~,
\eeq
and $ C_{i}^{\prime H} $ are given as
\beq
C_{1 \cdots 6}^{\prime \ H} (m_W)&=&0 ~,   \non
C_{7}^{\prime \ H} (m_W)&=&\frac{1}{2} \left [ \frac{m_s\
m_b}{m_t^2} \lambda_2^2 f_1 (y)+ \frac{m_s}{m_b} \lambda_1
\lambda_2 f_2 (y)+ \frac{m_s}{m_b} \lambda_1^2 f_1 (y) \right ]~,\non
C_{8}^{\prime \ H} (m_W)& = &\frac{1}{2} \left [
\frac{m_s \ m_b}{m_t^2} \lambda_2^2 g_1 (y)+ \frac{m_s}{m_b}
\lambda_1 \lambda_2 f_2 (y)+ \frac{m_s}{m_b} \lambda_1^2 g_1 (y)
\right ]~.
\eeq
In our calculations we neglect the
contributions of the internal $ u $ and $ c $ quarks and define
functions $f_1(y)$, $ f_2(y) $, $g_1(y)$, and $g_2 (y)$ as
\beq
f_1(y)&=& \frac{y (7- 5
y- 8 y^2)}{72 (y-1)^3} + \frac{y^2 (3 y -2)}{12 (y-1)^4} \ln[y]~, \non
f_2(y)&=& \frac{y (5 y- 3)}{12 (y-1)^2} + \frac{y
(-3 y +2)}{6 (y-1)^3} \ln[y] ~, \non
g_1(y)&=& \frac{y
(-y^2+5 y+2)}{24 (y-1)^3} + \frac{-y^2 }{4 (y-1)^4} \ln[y]~, \non
g_2(y)&=& \frac{y ( y- 3)}{4 (y-1)^2} + \frac{y)}{2
(y-1)^3} \ln[y]~,
\eeq
where $y = m_t^2/m_{H^{\pm}}^2$. Numerical analysis show that
$f_1(y)$ and $f_2(y)$ have the following properties:
\beq
f_1(y)<0,\ \ \ \ f_2(y)>0 ,\ \ \ \ \mid
f_2(y) \mid > 3 \mid f_1(y)~,  \mid \label{proper}
\eeq
and $g_i(y)$ have similar properties.

Using the renormalization group, we can get the LO Wilson coefficients at
any scale. The Wilson coefficients $ C_i^{III} (\mu) $ can be
obtained from \cite{bf97} with $ C_i^{SM}(m_W)$ replaced by $
C_i^{III} (m_W) $,  while $ C_i^{\, \prime \, III} (\mu) $ are
given by\footnote{Because the strong interaction preserve
chirality, the first set of operators (Eq.(\ref{operator})) cannot
mix with its chirality flipped counterparts, the primed operators,
and the anomalous dimension matrices of the two separate set of
operators are the same and do not overlap \cite{cm94}. This means
the evaluation of $C_i^{\prime} (\mu) $ are the same as in the
SM.}
\beq
C_{1 \cdots 6}^{ \prime \, III} (\mu)&= & 0 ~, \non
C_{7}^{\prime \, III} (\mu) &=& \eta^{\frac{16}{23}}
C_{7}^{\prime \, III } (m_W) + \frac{8}{3}
(\eta^{\frac{14}{23}}-\eta^{\frac{16}{23}} ) C_{8}^{\prime \,
III}(m_W)~,
\eeq
where $ \eta =\alpha_s (m_W) / \alpha_s (\mu) $.

The square amplitude summed over spin and polarizations can be
written as
\beq
|A|^2_{III}=|A|^2+|A^{\prime}|^2+2 |A^* A^{\prime} |~,
\eeq
where the first and  second term comes from $ O_1 \sim
O_8 $ and $ O_1^{\prime} \sim O_8^{\prime} $ respectively and the
third term is the interference between them. The explicit
expression of $|A|^2$ is the same as that in Eq.(\ref{eq:amps})
with $ C_i $ being replaced by $C_i^{III} $, and $ |A^{\prime} |^2
$ and $| A^* A^{\prime}|$ are given by
\beq
|A^{\prime}|^2&=&\frac{1}{4} \left (\frac{e^2 G_F}{\sqrt{2} \pi^2}
V_{t b} V_{ts}^* \right )^2 m_b^4|F^{\prime}_7|^2 A_{77}~,   \non
|A^* A^{\prime}|&=& \frac{1}{4} \left (\frac{e^2 G_F}{\sqrt{2}
\pi^2} V_{t b} V_{t s}^* \right )^2 m_b^4 \frac{m_s}{m_b} \left \{
|F_2^* F'_{7}| A'_{27}+|F_5^* F'_{7}| A'_{57}+|F_7^* F'_{7}|
A'_{77} \right \} \label{newamp}~,
\eeq
where
\beq
F^{\prime}_7 &=& C_7^{\prime \, III} Q_d ~,  \non
A'_{27} &=&-2*s-\frac{(1+\rho_1) s^2}{t (s-t)}+
s*(\rho_1-\rho)*(\frac{1}{t}-\frac{1}{s-t})+(t\leftrightarrow u)~,\non
A'_{57} &=&Re \left[8 (\kappa_b+\kappa_s) s- [4
(\kappa_b+ \rho_1 \kappa_s)+ 2 s (1-\kappa_b-\kappa_s) ]
\frac{s^2}{t (s-t)} \right. \non
& & \left. +4
(\kappa_b+\kappa_s) (\rho_1 -\rho) \frac{u}{s-t} \right ]
+(u\leftrightarrow t)~, \non
A'_{77}&=&2*\left [(1-\rho) A_{77}^1+A_{77}^2 \right ]-2 s \left ( 2+\frac{2}{t} +\frac{2}{u}
+\frac{2 (1- \rho_1-s)}{t u}  \right.  \non
&&\left.
-\frac{(\rho_1-\rho) (3+ 5 \rho_1) -(1+\rho_1) s}{(s-t)
(s-u)}-\frac{(1-\rho_1) (s^2+4 \rho_1 t u)}{(s-t) (s-u) t u}
\right )~.  \label{newamp1}
\eeq
In Eq.(\ref{newamp}-\ref{newamp1}), $ m_s $ is the current mass of $s
$ quark and the explicit expressions of $A_{77}, A_{77}^1 $ and $
A_{77}^2 $ have been given in Eqs.(\ref{eq:a77}-\ref{eq:a772}).
For $|A|^2_{III} $, two characters should be noted. Firstly, the
interference term between $O_i $ and $ O^{\prime}_i $ is
suppressed by $ m_s/m_b $ and therefore, to a good approximation,
can be neglected. Secondly, the total 1PR contributions are
proportional to $(|F_7|^2+ |F'_7|^2) A_{77}$. Since $|F_7|^2+
|F'_7|^2  $ is not suppressed, as required by $ B \to X_s \gamma $
in 2HDM \cite{ai99}, the 1PR contribution is then predominant,
which is very similar with the case in the SM. As we discussed in
section II, these characters mean that the  four differential
distributions defined in last section cannot have a large
deviation from SM predictions and we do not expect to see any new
physics signal by the measurement of these four differential
distributions. Indeed, we have checked our numerical results in a
vast parameter space constrained by Eq.(\ref{constr}), and find it
correct.

We are now ready to calculate the new physics corrections
to the decay rate of $ B \to X_s \gamma \gamma $.  In order to
reduce the effects of theoretical and experimental uncertainties,
we define two ratios
\beq
R_{\gamma \gamma} (\mu) &=& \frac{{\cal B} _ {B \to X_s \gamma \gamma}^{III}
(\mu)} {{\cal B} _ {B \to X_s \gamma \gamma}^{SM} (\mu)}~,\\
R_{\gamma} (\mu)  &=& \frac{{\cal B} _{B \to X_s \gamma }^{III}
(\mu)} {{\cal B} _{ B \to X_s \gamma}^{SM} (\mu) }~,
\eeq
where the four branching ratios denote the LO theoretical predictions
in the SM and model III, respectively.

In the numerical calculation, we consider the following three typical
scenarios.

\begin{itemize}

\item
{\bf Scenario I}:  $\ \ | \lambda_1 | = | \lambda_2 |=\lambda$ and $
m_{H^{\pm}} $ is about a few hundred GeVs.  This scenario can be
further divided into two limiting cases:

\begin{itemize}
\item
Case-1: $\lambda_1= \lambda_2 =\lambda $.

\item
Case-2: $\lambda_1=-\lambda_2=\lambda $ or $ \lambda_2=-\lambda_1=\lambda$,
where $ \lambda $ is positive.

\end{itemize}

\item
{\bf Scenario II}: $\lambda_2 =-2/\lambda_1=\lambda$ and $m_{H^{\pm}} $ is
about a few hundred GeVs.  By setting
\footnote{The $\lambda$ defined in this paper is different from that in
Ref.\cite{cck99} by a factor of $\sqrt{2}$.} $\lambda=-\sqrt{2}\, \tan{\beta}$,
one reproduces  the numerical results of the Model II, a very popular
version of the two-Higgs-doublet models.

\item
{\bf Scenario III}:  $ |\lambda_2 | \gg | \lambda_1 |$ and $
m_{H^{\pm}} $ is about $200$ GeV, which is favored by current
experiments.
\end{itemize}

Let us firstly concentrate on Scenario I. From the explicit
calculations, we find that the new physics contribution in case-1 tends to
increase the value of $C_7$. In Figs. \ref{fig7} and \ref{fig8}, we
show  the $\lambda$ dependence of the ratio $R_{\gamma \gamma}$ in
case-1 for fixed $m_{H^{\pm}} =300 GeV$ and assuming
$ \mu =2 m_b$ (dashed curve), $ m_b $ (solid curve) , $m_b/2 $
(dotted curve), respectively.
In determining the range of $ \lambda $, we have
put the constraint of Eq.(\ref{constr}) and required $ C_7 $ to be
negative and positive respectively.

For the case-1, one can see  from Figs. \ref{fig7} and \ref{fig8} that
\begin{itemize}
\item
The $\mu$ dependence of the ratio $\rgg$ is weak for a negative
$C_7$: we have $0.7\leq \rgg \leq 1.0 $ within the range of $1/2 \leq \mu/m_b \leq 2$.

\item
The $\mu$ dependence of the ratio $\rgg$ is strong for a positive $C_7$:
we have $0.3\leq \rgg \leq 2.7 $ within the range of $1/2 \leq \mu/m_b \leq 2$;
which implies that the new physics contribution to the branching ratio $\bsgg$
can be significant.

\item
In the same region of $\lambda$ as specified in Figs. \ref{fig7} and
\ref{fig8}, the numerical results show that the ratio $\rgg / \rg $  only varies
within  a very small range of $ 1\pm 0.02$ for $ 1/2 \leq \mu/m_b \leq 2 $.
This fact reflects the dominance of 1PR contribution to the decay rate of
$ \bsgg$, and a strong correlation between the branching ratios of
$ \bsgg$ and $ \bsga $.

\end{itemize}

The main feature of case-2 is that the new physics contribution
tends to decrease  the value of $C_7$,  as a result, coefficient
$C_7$ keeps negative for all the values of $\lambda$.
In Fig. \ref{fig9}, we
show the dependence of $\rgg $ on $ \lambda $ in
case-2 for fixed $m_{H^{\pm}} =300 GeV$ and $ \mu =2 m_b $ (dashed
curve) , $ m_b $ (solid curve), $m_b/2 $ (dotted curve). It is easy to see
that the $\mu$ dependence of $R_{\gamma \gamma}$ is weak. Study of
the ratio $ R_{\gamma \gamma} / R_{\gamma} $ gives the same conclusion
as in case-1.

In Fig. \ref{fig10},  we plot the $ m_{H^{\pm}}$
dependence of the ratio $ R_{\gamma \gamma} $ in Scenario I.
As one can see from this figure, $R_{\gamma\gamma}$ tends to
approach 1 when the charged-Higgs boson becomes heavier, which is
just the decoupling behaviour of the new heavy particle.

Secondly, we consider the Scenario II. By setting $\lambda=-\sqrt{2}
\tan{\beta}$, one reproduces the result of the popular model II: the second type of
the two-Higgs-doublet models. Wilson coefficient $ C_7 $ keeps negative for all
the values of $\lambda$. In model II, two additional free parameters will enter into
our calculation: the charged Higgs mass $\mhp$ and the ratio $\tan{\beta}=v_2/v_1$
with $v_2$ and $v_1$ are the vacuum expectation value of the two Higgs doublets
$\phi_1$ and $\phi_2$.

In Fig.\ref{fig11}, we plot the $\tan{\beta}$ and $\mu$ dependences of the ratio
$R_{\gamma \gamma}$ in model II for fixed $m_{H^{\pm}} =500 GeV$.
In this figure, the three curves correspond to $ \mu
=2 m_b$ (dashed curve), $ m_b $ (solid curve), and $m_b/2 $ (dotted
curve), respectively. One can see that (a) the ratio $\rgg$ is larger than $1.32$
in the whole considered range of $\tan{\beta}$ and $\mu$, and (b)
the ratio $\rgg$ has a weak $\tan{\beta}$ dependence for $\tan{\beta} \geq 3$,
but sensitive to the variation of scale $\mu$.

In Fig.\ref{fig12}, we plot the $\tan{\beta}$ and $\mhp$ dependences of the ratio
$R_{\gamma \gamma}$ in model II for fixed $\mu=m_b$.
In this figure, the three curves correspond to $ \tan{\beta}=1
$ (solid curve), $ 10$ (dashed curve), and $100$ (dotted
curve), respectively. One can see that the ratio $\rgg$ is insensitive to
the variation of $\tan{\beta}$.

Finally, we consider the Scenario III. The main characteristic
of this Scenario is that $C_7$ increases with increasing $ \lambda_1$
for fixed $\lambda_2 $.
In Figs. \ref{fig13} and \ref{fig14}, we illustrate the $\lambda_1$ and $\mu$
dependences of the ratio $R_{\gamma \gamma}$ for fixed $\mhp=200$ GeV and
$ \lambda_2=50$. In these two figures, the three curves correspond to
$ \mu =2 m_b$(dashed curve), $m_b$ (solid curve), and $m_b/2$ (dotted curve),
respectively. In determining the range of $ \lambda_1 $, we have considered
the constraint of Eq.(\ref{constr}) and required $ C_7 $ to be negative and
positive for Figs. \ref{fig13} and \ref{fig14}, respectively.
Again,  the ratio $ R_{\gamma \gamma} / R_{\gamma}$ only varies within a
very small range of $ 1\pm 0.02$ for $ 1/2 \leq \mu/m_b \leq 2 $.

In Fig. \ref{fig15}, we show the $\lambda_2$ and $\mu$
dependences of the ratio $R_{\gamma \gamma}$ for fixed $\mhp=200$ GeV and
$ \lambda_1=0.2$. In this figure, the three curves correspond to
$ \mu =2 m_b$(dashed curve), $m_b$ (solid curve), and $m_b/2$ (dotted curve),
respectively. In determining the range of $ \lambda_2 $, we have required
Eq.(\ref{constr}) to be satisfied. It is easy to see that the ratio $\rgg$
has a strong (moderate) dependence on $\mu$ ($\lambda_2$). The future NLO
calculation of $\rgg$ in the SM and 2HDM's  will decrease the $\mu$ dependence.
Study of the ratio $R_{\gamma \gamma}/ R_{\gamma} $ also shows strong correlations
between the branching ratios of $\bsga$ and $\bsgg$ decays.

In Figs. \ref{fig16} and \ref{fig17}, we plot $\mhp$, $\lambda_1$ and $\lambda_2$
dependences of the ratio $\rgg$ in model III and assuming $\mu=m_b$.
In Fig. \ref{fig16},
the soild and dashed curve refers to $\lambda_1=-0.018$ and $0.010$,
respectively. When choosing the value of $\lambda_1$, we keep $C_7$ to be
negative and require  $ |C_7 | >|C_7^{SM}| $ ( solid curve) and
$ |C_7 |< |C_7^{SM}| $ (dashed curve), respectively.
One can find from Fig. \ref{fig16} that the ratio $\rgg$ tends to approach 1
when charged Higgs boson is becoming heavier.
In Fig. \ref{fig17}, the three curves correspond to $\lambda_1=0.2$,
$\lambda_2=40$(dashed curve), $50$ (soild curve), and $60$ (dotted curve), respectively.
To determine  the range of $\mhp$, we have required  Eq.(\ref{constr}) to be satisfied
and $C_7 > 0 $. Numerical  results also show  that for a much larger $\mhp$,
$C_7 $ is driven to be negative.

\section{conclusion}

In this paper, based on the low-energy effective Hamiltonian, we
calculated the rare decay $B \to X_s \gamma \gamma$ in the SM and
the general two-Higgs-doublet model with the restriction
$\lambda_{ij}^{U,D}=0$ for $i\neq j$.
We focused on the estimation of the new physics
contributions to the branching ratio ${\cal B}(B \to X_s \gamma
\gamma)$ and the differential distributions $(1/\Gamma)
d\Gamma/ds$, $(1/\Gamma) d\Gamma/d\cos \theta_{\gamma \gamma}$ and
$(1/\Gamma) d\Gamma/d X_{\gamma}$.

Within the considered parameter space allowed by currently
available data, we found the following:
\begin{itemize}

\item
In model III, the prediction  of the branching ratio $ {\cal B}( B
\to X_s \gamma \gamma) $ ranges  from one third  to
three times of the SM prediction, but it is highly correlated with the
corresponding theoretical prediction of $ {\cal B}( B \to X_s \gamma )$.

\item
In model II, the new physics enhancement to the branching ratio $ {\cal B}( B
\to X_s \gamma \gamma) $ can be as large as $(30-50)\%$ with respect to the
SM prediction.

\item
In the SM and model III, the contribution  from 1PR diagrams
is dominant and hence those four observable differential
distributions are insensitive to the variation of scale $\mu $  as
well as possible new physics corrections considered in this paper.

\item
Although the process $ B \to X_s \gamma \gamma $ provides many new physical
observables,  it is not a better process to detect new physics than
$ B \to X_s \gamma $ because of the smallness of its decay rate and the
long-distance background.

\end{itemize}

At the next-to-leading order, the coefficients of $O_3  - O_6 $ and
$ O_1^{\prime}  -  O_6^{\prime} $ may get enhanced. However, due to their subleading
feature, we do not expect drastic deviation from the LO predictions.

\section*{ACKNOWLEDGMENTS}

The authors acknowledge the support by the National
Natural Science Foundation of China under Grant Nos. 19575015, 19775012 and
10075013, and the Excellent Young Teachers Program of Ministry of Education,
P.R.China.



\newpage

\begin{table}
\begin{center}
\caption{Values of the regularization-scheme-independent LO Wilson
coefficients  $C_i(\mu)$ using  the input parameters as listed in Table II. }
\label{table1}
\begin{tabular}{c c c c c c c c}
 $\mu $ &  $C_1 $  & $C_2 $  & $C_3$ & $C_4$ & $C_5 $ & $C_6 $ & $ C_7$ \\  \hline
$m_b /2 $  & $-0.315 $ & $1.143  $ & $0.015 $ & $ -0.032$ & $0.009
$& $-0.041 $& $ -0.340$
 \\ \hline
$m_b $  & $-0.225 $ & $1.10  $ & $0.010 $ & $ -0.023$ & $0.007 $&
$-0.028 $& $ -0.304 $
 \\ \hline
$2 m_b $  & $-0.154 $ & $1.061  $ & $0.007 $ & $ -0.016$ & $0.005
$& $-0.019 $& $ -0.273$
\end{tabular} \end{center}
\end{table}

\begin{table}
\begin{center}
\caption{Values of the input parameters used in the numerical
calculations.}
\label{table2}
\begin{tabular}{c c c c c c c }
 $ \alpha_s (m_Z)$ &  $\alpha_e $ & $m_t $  & $m_Z$ & $m_W$ & $m_b $ & $m_c $ \\  \hline
$ 0.118 $  & $ 1/129 $ & $175GeV  $ & $91.2GeV $ & $ 80.4 GeV$ &
$4.8 GeV $& $ 1.5 GeV $  \\ \hline $m_s (current) $ & $m_K $  &
$m_u $ & $m_d $ & $ |V_{t s}^* V_{t b} | $ & $ |V_{b c}|$ & ${\cal B} (b
\to X_c e \upsilon_e) $  \\ \hline
$0.15 GeV $  & $0.5 GeV $ & $5.1 MeV  $ & $9.0 MeV$ & $ 0.04$ &
$0.04 $& $0.11 $
\end{tabular} \end{center}
\end{table}

\begin{table}
\begin{center}
\caption{Current constrains on
$\lambda_{ij}^{U,D} $ along with the processes from which
constrains have been or will be placed. Couplings that do not
appear in this table are not constrained. }
\label{table3}
\begin{tabular}{c c c c}
 coupling &  current constrain  &  current
constrain  from  & future constrain from \\  \hline
 $ \lambda_{s d}^D$ & $\ll 1 $ & $ K^0-\bar{K}^0 $  mixing &  \\ \hline
$ \lambda_{b d}^D $ & $\ll 1 $ & $ B_d-\bar{B}_d^0 $ mixing &  \\\hline
$ \lambda_{u c}^U $ & $ \ll 1 $ &  $D^0-\bar{D}^0 $ mixing & \\ \hline
$ \lambda_{t t}^U $ &  $<0.5 $ & $ R_b $, $ B_d-\bar{B}_d^0 $ mixing &   \\ \hline
$ \lambda_{b b}^D $ & $ \geq 40 $ & $ R_b $, $ R_c $ & \\  \hline
$ \lambda_{s b}^D $ & $ \leq 40 $ & $ b \to s c c $ and
$ B_s-\bar{B}_s   $ & $ Z \to b s $, $B_{d, s}^0 \to l^+ l^- $, $ b \to X_s \mu^+ \mu^- $ \\  \hline
$ \lambda_{t c}^u $ &  - & - & $ e^+ e^- \to t c$, $ t \to c \gamma, c Z , c g$
\end{tabular} \end{center} \end{table}

\end{document}